\documentclass{aa}  

\usepackage{graphicx}
\usepackage{multirow}   %multirows in tables
\usepackage{lscape}     %for table landscape format in appendix
\usepackage{txfonts}
\begin{document}

   \title{Search for flares and associated CMEs on late-type main-sequence stars in optical SDSS spectra \protect\thanks{The full tables containing the flare analysis results are only available in electronic form
at the CDS via anonymous ftp to cdsarc.u-strasbg.fr (130.79.128.5)
or via http://cdsweb.u-strasbg.fr/cgi-bin/qcat?J/A+A/}}

  % \subtitle{I. Overviewing the $\kappa$-mechanism}

\author{Florian Koller 
\inst{1}
\and
Martin Leitzinger \inst{1} \and
Manuela Temmer \inst{1} \and
Petra Odert \inst{1} \and
Paul G. Beck \inst{1,2} \and
Astrid Veronig\inst{1}
}

\institute{Institute of Physics/IGAM, University of Graz, Universitätsplatz 5, A-8010 Graz, Austria
\and Instituto de Astrof\'{\i}sica de Canarias, E-38200 La Laguna, Tenerife, Spain
%\and Departamento de Astrof\'{\i}sica, Universidad de La Laguna, E-38206 La Laguna, Tenerife, Spain
}

   \date{Received xxx; accepted yyy}

% \abstract{}{}{}{}{} 
% 5 {} token are mandatory
 
  \abstract
   {}
   {This work aims to detect and classify stellar flares and potential stellar coronal mass ejection (CME) signatures in optical spectra provided by the Sloan Digital Sky Survey (SDSS) data release 14. The sample is constrained to all F, G, K, and M main-sequence type stars, resulting in more than 630,000 stars. This work makes use of the individual spectral exposures provided by the SDSS.}
   {An automatic flare search was performed by detecting significant amplitude changes in the $H\alpha$ and $H\beta$ spectral lines after a Gaussian profile was fit to the line core. CMEs were searched for by identifying asymmetries in the Balmer lines caused by the Doppler effect of plasma motions in the line of sight.}
   {We identified 281 flares on late-type stars (spectral types K3 – M9). We identified six possible CME candidates showing excess flux in Balmer line wings. Flare energies in $H\alpha$ were calculated and masses of the CME candidates were estimated. The derived $H\alpha$ flare energies range from $3 \times 10^{28} - 2 \times 10^{33}$ erg. The $H\alpha$ flare energy increases with earlier types, while the fraction of flaring times increases with later types. Mass estimates for the CME candidates are in the range of $6 \times 10^{16} - 6 \times 10^{18}$ g, and the highest projected velocities are $\sim300 - 700\:$ km/s.}
   {The low detection rate of CMEs we obtained agrees with previous studies, suggesting that for late-type main-sequence stars the CME occurrence rate that can be detected with optical spectroscopy is low.}


   \maketitle
%
%-------------------------------------------------------------------

%%%%%%%%%%%%%%Introduction%%%%%%%%%%%%%%%%%%%%
\section{Introduction}\label{sec:intro}
The activity of the present-day Sun is investigated in detail, but an ambiguity remains about how active the Sun was in its earlier life. This varying activity strongly affects the evolution of the planetary atmospheres in the Solar System. Observing solar-like stars with different ages would therefore give us an idea of the Sun in time, meaning the evolution of solar-like stars throughout their lifespan \citep{sonett_sun_in_time_1991suti.book.....S, ribas_evolution_solar_activity_2005ApJ...622..680R,guedel_sun_in_time_2007LRSP....4....3G}. 

While solar flares are frequently observed, they would hardly be detectable using optical disk-integrated measurements  \citep{tayler_suns_as_star_1997sas..book.....T,kretzschmar_sun_as_star_flares_2011A&A...530A..84K,pevtsov_sun_as_a_star_spectroscopy_2014AN....335...21P}. Highly energetic stellar flares on late-type stars have been observed several times, with later types (M dwarfs) showing more flaring activity than earlier types  \citep{pettersen_review_of_flares_in_hrd_1989SoPh..121..299P,balona_flares_across_hrd_kepler_2015MNRAS.447.2714B,van_doorsselaere_flares_kepler_2017ApJS..232...26V}. Highly energetic flares on solar-type stars have also been reported, however 
\citep[bolometric energy of $ 10^{33} - 10^{36}\: \mathrm{erg}$, up to four orders of magnitude higher than the largest solar flares,
see][]{maehara_superflares_solartypes_2012Natur.485..478M,BASRI_solar_stellar_connection_2019363}.
While solar coronal mass ejections (CMEs) have also frequently been detected in the past decades with several CMEs per day during high activity \citep[e.g.][]{webb_cmes_observation_sun_review_2012LRSP....9....3W}, reported CME detections on other stars have so far been rare. On the Sun, the association between highly energetic flares and CMEs is strong and reaches almost $100\:\%$ for the strongest flares \citep{yashiro_association_cme_flares_2006ApJ...650L.143Y}. If the same association rate of solar CMEs to  flare were assumed to hold on stars, the habitability of exoplanets would be severely affected. 
It would also lead to a stellar mass-loss rate that would exceed so far observed total mass-loss rates on highly active stars, implying that the strong correlation between energetic flares and CMEs might be more uncertain on these objects \citep{drake_mass_loss_due_to_cme_stellar_2013ApJ...764..170D,odert_stellar_cmes_occurence_mass_loss_rates_2017MNRAS.472..876O}. 

Because direct imaging of stellar CMEs is not feasible with current instrumentation, the detection of CMEs is based on other methods. CMEs that have been reported so far were mainly found using signatures such as Doppler shifts of spectral lines and X-ray absorptions \citep{stellar_cme_flare_relation_moschou_2019ApJ...877..105M}. 

The Doppler-shift method uses enhancements in the blue or red wings in spectral lines (in particular the Balmer lines), implying that this flux comes from ejected material moving toward or away from the observer along the line of sight. This means that the velocities determined by this method are projected velocities giving a lower estimate. Only the erupting filament or prominence, which builds the dense cores of CMEs, is visible as a Doppler feature. The first CME reported with this method was presented by \citet{houdebine_cme_1990A&A...238..249H}, showing a large blue wing enhancement in $H\gamma$ with a projected maximum velocity up to $5800\: \mathrm{km\:s^{-1}}$. \citet{guenther_cme_on_wtts_1997A&A...321..803G} reported a CME detection with a bulk velocity of $\sim 600\: \mathrm{km\:s^{-1}}$. Several other blueshifted emissions have been reported, possibly arising from a CME \citep{gunn_high_vel_chrom_evaportaion1994A&A...285..489G,fuhrmeister_cme_on_m9_2004A&A...420.1079F,fuv_cme_search_leitzinger_2011A&A...536A..62L,vida_v374peg_cme_activity_2016A&A...590A..11V}. 
The reported velocities of these studies range from $\sim 100\:\mathrm{km\:s^{-1}}$ up to several thousand $\mathrm{km\:s^{-1}}$; the event studied by \citet{houdebine_cme_1990A&A...238..249H} was the fastest one. The estimated CME masses from these reports range from $10^{15}\:\mathrm{g}$ up to $10^{19}\:\mathrm{g}$. \citet{vida_quest_for_stellar_cmes_2019A&A...623A..49V} searched for asymmetries in Balmer lines in both wings on near M-dwarfs, finding 478 (9 strong) asymmetries in more than 5500 spectra. They reported projected velocities ranging from $100$ to $300\:\mathrm{km\:s^{-1}}$ and estimated masses of $10^{15} - 10^{18}\:\mathrm{g}$, concluding that their reported maximum velocities are lower than solar CMEs \citep[with typical velocities of about $500\:\mathrm{km\:s^{-1}}$,][]{gopalswamy_statistic_cmes_sun_2010ASSP...19..289G} and a somewhat lower event rate, suggesting that CMEs might be suppressed by a strong overlying magnetic field. Observing at shorter wavelengths, \citet{argiroffi_stellar_flare_cme_2019NatAs...3..742A} reported a stellar flare event including a possible CME using X-ray measurements, with Doppler-shifted plasma movements during the flare followed by a blueshifted upward motion of $\sim 90\:\mathrm{km\:s^{-1}}$ of cool plasma. Several studies have also searched for CMEs without yielding any detections \citep{leitzinger_flare_cme_search_blanco_2014MNRAS.443..898L, korhonen_hunting_cmes_2017IAUS..328..198K, leitzinger_census_cmes_solar_like_2020MNRAS.493.4570L}. CMEs may produce an absorption feature if the erupting filament lies directly on the line of sight to the observer. \citet{den_cme_solar_filtergrams1993ARep...37...76D} and
\citet{ding_solar_filament_eruption_blue_shift_2003ApJ...598..683D} analyzed an erupting filament on the Sun that was observed as a blue wing absorption in $H\alpha$, while also noting that some parts of the material showed emission instead of absorption. No absorption feature reminiscent of an erupting prominence has so far been reported on stars other than the Sun.

Another method to detect CMEs is based on space-borne measurements of X-ray absorptions. The increase in the absorbing column density is interpreted as a rising CME above an active region. Such X-ray absorption events were reported, for example, by \citet{favata_cme_xray_abs_algol_1999A&A...350..900F},  \citet{haisch_xray_cme_proximacen_1983ApJ...267..280H}, and \citet{moschou_moster_cme_2017ApJ...850..191M}. In the extreme UV (EUV) range, we can search for EUV dimmings caused by a CME because plasma is evacuated during the beginning phase. Dimmings associated with CMEs are frequently observed on the Sun \citep[see e.g.][]{coronal_dimming_dissauer_2018ApJ...855..137D}. \citet{ambruster_1986ESASP.263..137A} reported evidence for a stellar case of a UV dimming (they interpreted it as obscuring CME matter). Type II bursts in the radio domain are another potential method for identifying CMEs. Type II bursts are a signature of electrons accelerated at shock fronts driven ahead of fast CMEs, but no stellar type II bursts have been detected so far \citep{leitzinger_decametric_obs_m_stars_2009AIPC.1094..680L,boiko_radio_typeii_2012AASP....2..121B,crosley_typeiibursts_2018ApJ...856...39C,crosley_typeiibursts_2018ApJ...862..113C,villadsen_typeii_bursts_2019ApJ...871..214V,mullan_cmes_radio_quiet_2019ApJ...873....1M}.
We refer to \cite{stellar_cme_flare_relation_moschou_2019ApJ...877..105M} and references therein for a more complete overview on CME reports in the literature. 

The searches for flares and CMEs have in common that stellar observations over long time periods are needed in order to detect these events. A promising approach are time-resolved observations of stars in data archives, with which the long observation time necessary for detecting such sporadic stellar activity phenomena can be achieved. The problem of having only snapshots instead of long, consecutive observations of single objects might be compensated by having a large number of objects in an archive.

The aim of this work is to search for flares and CMEs on late-type main-sequence stars (F - M) using optical spectra of the Sloan Digital Sky Survey (SDSS). Most of the stars in the SDSS are observed several times in order to obtain a coadded spectrum with better signal-to-noise ratio (S/N) than the single spectra. However, these single observations can be used to find temporal changes in the Balmer lines that indicate possible activity in the chromosphere. Only a few studies have used time-resolved optical SDSS spectra to infer stellar activity \citep{kruse_sdss_chrom_variability_time_resolved_spectra_2010ApJ...722.1352K,hilton_2010AJ....140.1402H,bell_emission_variable_m_dwarfs_sdss_spectra_2012PASP..124...14B}. \citet{hilton_2010AJ....140.1402H} searched for flares on M dwarfs in optical SDSS spectroscopic data. 
With respect to these previous publications, we extend the dataset, use a different method, a different sample selection, extend the stellar subtypes to F, G, and K stars, perform searches for line asymmetries and CMEs, and use the latest GAIA DR2 data to derive stellar distances, based on which we calculate absolute values of $H\alpha$ flare energies and mass estimates for CME candidates.

%%%%%%%%%%%%%%%%DATA%%%%%%%%%%%%%%%%
\section{Data sample}\label{sec:data}

\subsection{SDSS spectroscopic data} \label{subsec:sdss_spectroscopic_data_overview}

This work makes use of the optical spectroscopic data of the SDSS data release 14 \citep{sdss_dr14_2018}. The spectra were restricted to all stars of type F, G, K, and M. 
Overall, 630,162 spectra were available for the analysis. Stars that are specifically marked with luminosity class I, II, and III as well as white dwarf stars were excluded. Every spectrum is a combination (called the coadded spectrum) of one or more single spectra. These single spectra were usually observed within one hour to up to several weeks. Only objects with at least three single spectrum observations were used to obtain information on the time evolution. It is possible that one object has several coadded spectra, each comprised of a number of single spectra (this usually happens when an object was observed in different survey programs). The exposure times of the single spectra are in the range of roughly 10 to 25 minutes. Each of the single spectra itself is split into a red and a blue part. The spectra are provided in units of spectral flux density $[ 10^{-17}\:\mathrm{erg \: s^{-1} cm^{-2} \text{\AA} ^{-1}} ],$ and the wavelength axis is given in vacuum wavelengths in the heliocentric frame \citep{sdss_early_data_release_flags_pipeline2002AJ....123..485S}.

\begin{figure*}
 \centering
 \includegraphics[width=17cm]{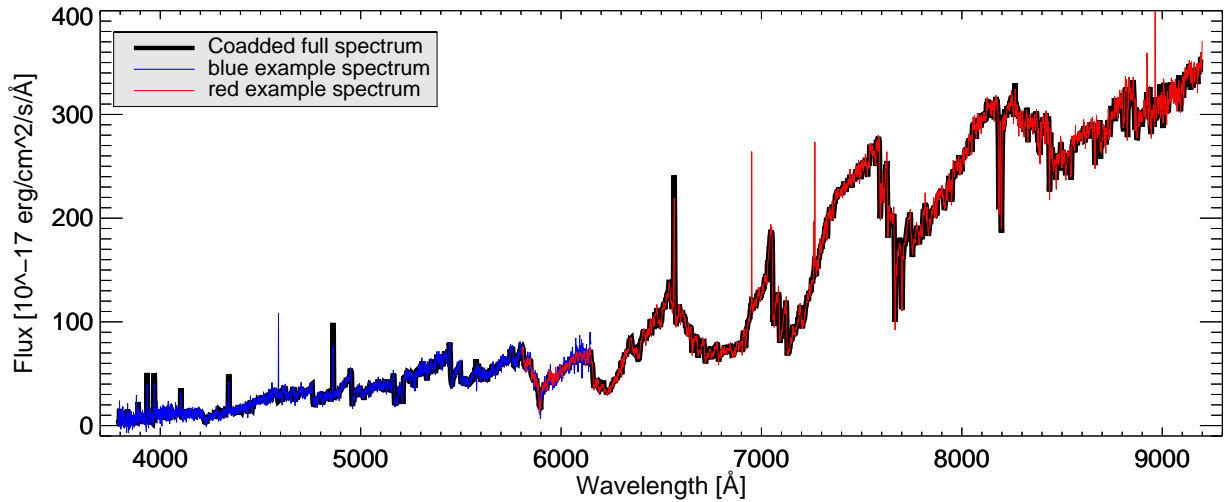}
 \caption{Full SDSS optical range example plot of a magnetically active M5e star. The coadded spectrum (black) consists of several spectra that are divided into a red and a blue part, overlapping at $\sim 6000\:\text{\AA}$. One blue and one red single spectrum is plotted. Cosmic rays in the single spectra are visible as narrow peaks (one in the blue and up to four in the red spectrum range). \label{fig:example_sdss}}
 \end{figure*}

The SDSS data release 14 includes all the data from previous data releases, covering a wide range of different surveys and programs. Two spectrographs were used: the SDSS spectrograph covering data releases 1 - 8, including the original SDSS spectroscopical survey as well as the Sloan Extension for Galactic Understanding and Exploration (SEGUE) surveys, and BOSS (Baryon Oscillation Spectroscopic Survey) starting with data release 9. The SDSS observes many targets at once using a plate with predrilled holes for each object in the observed sky patch. The original spectrograph was able to observe 640 objects at the same time with a wavelength range of $3800 - 9200 \: \mathrm{\text{\AA}}$. The BOSS spectrograph is able to observe 1000 objects at once with a slightly wider wavelength range of $3650 - 10400\: \mathrm{\text{\AA}}$. Both give a spectral resolution of 1500 at $3800\: \mathrm{\text{\AA}}$  and 2500 at $9000\: \mathrm{\text{\AA}}$. The split between the red and blue part of the optical data is at about $6000\: \mathrm{\text{\AA}}$ \citep{sdss_dr9_with_boss_2012ApJS..203...21A}. One example of such a spectrum is displayed in Fig. \ref{fig:example_sdss}.

Because more stars are observed simultaneously, the runtime of surveys is long, and the datasets are cumulative, the overall summed observation time of the stars in our dataset is about 131 years. All pure exposure times of the files were summed and are listed in Table \ref{tab:obs_times}.

Flux errors in the spectra are given as estimated $1 \sigma$ standard deviations per pixel, assuming a normal distribution for each pixel. The SDSS data pipeline slightly adjusted the spectra to match telluric lines and to correct to the heliocentric frame, reporting an accuracy of the wavelength calibration of $10\: \mathrm{km\:s^{-1}}$ or better \citep{sdss_early_data_release_flags_pipeline2002AJ....123..485S}.

This work focuses on the analysis of the $H\alpha$ and $H\beta$ lines because the S/N is better than at higher Balmer lines. With a given approximated spectral resolving power R = 2000 within the two Balmer line ranges, we obtain a  spectral resolution of at least $\sim 3.3\:\mathrm{\text{\AA}} $ around $H\alpha$ and $\sim2.4\: \text{\AA}$ around $H\beta$ (approximately twice oversampled in both ranges).

\begin{table*}
\caption{Summed exposure times of all stars in the dataset.}
\centering
\begin{tabular}{l|cccc|c} \hline \hline
\multicolumn{1}{l|}{\multirow{2}{*}{\begin{tabular}[l]{@{}l@{}}Stellar \\ spectral type\end{tabular}}} & \multicolumn{4}{c|}{Summed exposure times per S/N bin {[}years{]}} & \multirow{2}{*}{\begin{tabular}[c]{@{}c@{}}Sum \\ {[}years{]}\end{tabular}} \\
\multicolumn{1}{c|}{} & High S/N & Medium S/N & Low S/N & \multicolumn{1}{l|}{Lowest S/N} &  \\ \hline
M & 1.39 & 1.69 & 2.11 & 12.56 & 17.75 \\
K & 13.86 & 4.97 & 3.67 & 9.40 & 31.9 \\
G & 4.75 & 2.26 & 2.61 & 6.70 & 16.32 \\
F & 22.55 & 10.36 & 12.16 & 20.10 & 65.17 \\ \hline
Sum & 42.55 & 19.28 & 20.55 & 48.76 & 131.14\\ \hline
\end{tabular}\tablefoot{Dataset is divided into stellar spectral types and S/N bins. The cutoff values defining each bin are given in Sect. \ref{subsec:snrbins}} \label{tab:obs_times}
\end{table*}

\subsection{Spectral classification}\label{subsec:spectral}
Both spectrographs used specific spectral templates for each stellar subtype, taking the template closest to the spectrum at hand to classify the star. While the original SDSS legacy surveys and the SEGUE had templates for dwarf stars of subtype of M0 - M9, BOSS only used a template for dwarf stars of subtype M1 in the same classification region, leading to fewer late-type M stars since SDSS data release 9. It is noted that SDSS spectral classifications often do not explicitly state the luminosity class.

The spectral type of every star that was found by the stellar activity search was confirmed using \citet{spectral_class_2014yCat....1.2023S} on VizieR, which is a collection of known spectral classifications of different literature sources for each star. One of these sources is \citet{west_m_dwarfs_sdss_spectral_types_distances_2011AJ....141...97W}, a catalog of the SDSS M dwarfs up to data release 7. Most of our stellar activity objects were matched with this catalog. \citet{west_m_dwarfs_sdss_spectral_types_distances_2011AJ....141...97W} performed visual inspection of each object to determine the spectral type. The same catalog was also frequently used to obtain distances when an object was not in GAIA.

\subsection{S/N bins}\label{subsec:snrbins}
\begin{figure*}
 \centering
 \includegraphics[width=17cm]{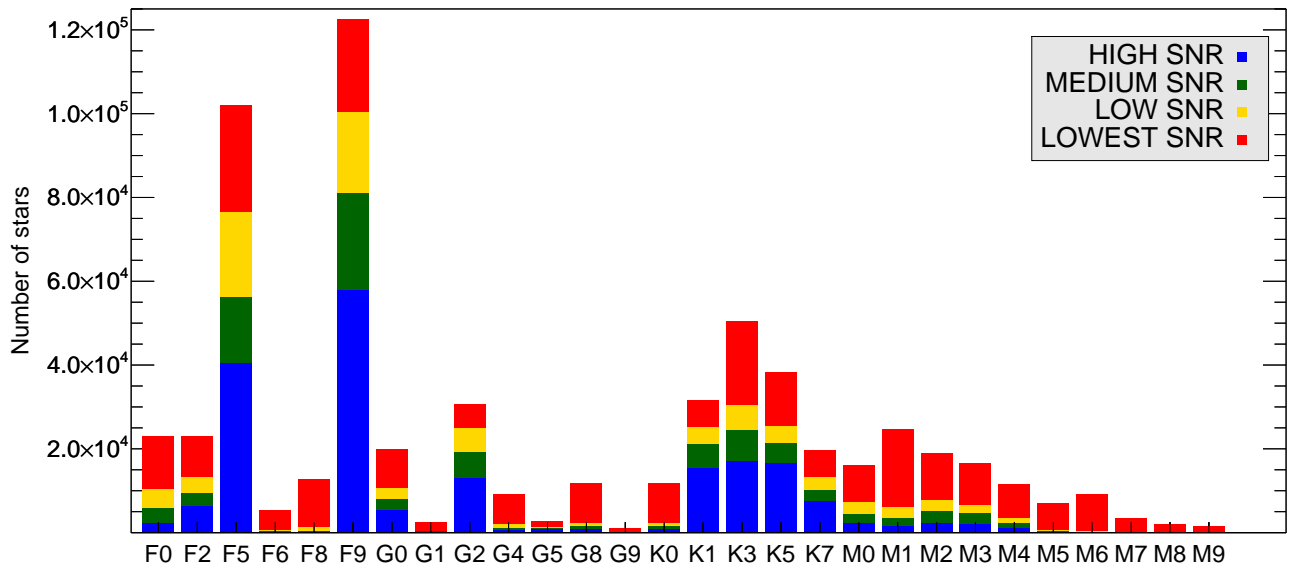}
 \caption{Number of stars per stellar subtypes in each S/N bin. S/N bins are stacked in this visualization. Spectral classification according to the SDSS.
 \label{fig:subtypes_snrbins_all}}
 \end{figure*}

All spectra were classified into different S/N bins by calculating the S/N within +/- $200\: \mathrm{\text{\AA}}$ around the $H\alpha$ line. The reason for the classification based on the surroundings of the $H\alpha$ line is that its continuum flux on late-type stars is higher than for the other Balmer lines. For each object, the mean S/N of all single spectra was calculated. This led to the following classification:
\begin{itemize}
    \item High-S/N bin: S/N around $H\alpha$ $\geq 20,$ resulting in 197,047 stars ($31.3\:\%$).
    \item Medium-S/N bin: S/N around $H\alpha$ between $< 20$ and $\geq 15,$ resulting in 88,144 stars ($14.0\:\%$).
    \item Low-S/N bin: S/N around $H\alpha$ between $< 15$ and $\geq 10,$ resulting in 90,119 stars ($14.3\:\%$).
    \item Lowest S/N bin: S/N around $H\alpha$ $\leq 10,$ resulting in 254,852 stars ($40.4\:\%$).
\end{itemize}
Figure \ref{fig:subtypes_snrbins_all} shows the number of stars according to the SDSS spectral subtype classification for each S/N bin. As was mentioned in Sect. \ref{subsec:spectral}, this classification is dependent on the spectral templates used by SDSS. Since BOSS changed the spectral templates, some subtypes are either over- or underrepresented in the data. Figure \ref{fig:subtypes_snrbins_all} shows that the later spectral subtypes are mostly prevalent with poor S/N, as expected because their luminosities are low.

\subsection{Cosmic rays and flags}\label{subsec:res,crs,flags}
Cosmic rays are visible in the data, usually compromising two neighboring data points. Every spectrum found by our algorithms was analyzed by eye to rule out that cosmic rays result in false positives in the stellar activity search. The SDSS provides flags for each data point, which provides information on occurring problems. These flags are resourceful in noting large cosmic rays as well as spectra with general problems, while on the other hand, stellar activity is often also flagged (due to sudden changes from one spectrum to another, as well as large differences to the SDSS template spectrum). We therefore considered flags and ruled out cosmic rays after the first search for stellar activity because the SDSS pipeline does not differentiate between cosmic rays and stellar activity. More background information on the pipelines, calibration, and flags of the spectroscopic data can be found in \citet{sdss_early_data_release_flags_pipeline2002AJ....123..485S}.

\subsection{GAIA DR2 data}\label{subsec:gaiadr2}
We used the second GAIA data release \footnote{\label{gaia} \url{http://cdsweb.u-strasbg.fr/gaia}}. GAIA DR2 gives information on position and apparent G magnitudes for approximately 1.7 billion sources, parallaxes of 1.3 billion sources, and apparent red photometer (RP) and blue photometer (BP) magnitudes for 1.4 billion sources \citep{2018gaiadr2}. Only sources brighter than 21 magnitudes are considered by GAIA.

To match each of the 630,162 spectroscopic sources from SDSS with their GAIA DR2 counterparts, a VizieR search within the GAIA DR2 catalog was conducted for the position of every single SDSS source. The search radius was 4 arcseconds. GAIA G magnitude information is available for approximately 591,000 stars in our sample ($\sim 94\:\%$), while BP-RP color information is available for $\sim 90\:\%$ of the stars in the sample. Both GAIA G magnitudes and parallaxes are available for approximately 480,000 stars ($\sim 76\:\%$).

The distances were calculated using the simple relation of inverting the GAIA DR2 parallax (in units of arcseconds) to derive the distance (in units of parsecs). There are more sophisticated methods to derive the correct distance from GAIA DR2 data \citep[see][]{bailer-jones_distances_gaia_dr22018AJ....156...58B}, but the resulting difference in distance values between our work and \citet{bailer-jones_distances_gaia_dr22018AJ....156...58B} lies within the parallax errors provided by GAIA DR2.
While the cross-matching of our objects with the GAIA DR2 catalog is valid for the majority of the stars, a few special cases remained that were treated separately (e.g., the possibility of two GAIA sources within 4 arcseconds, see Appendix \ref{sec:appendix_special_cases_flares} for examples of such cases). 
\subsection{Data preparation}\label{subsec:dataprep}

The methods we used to search for flares and CMEs are based on fitting algorithms of Balmer lines. To correctly perform these fits, the spectroscopic lines were normalized. Spectral ranges for lines and continua were adopted from \citet{hilton_2010AJ....140.1402H} and are listed in Table \ref{tab:normalizatio_ranges}. For each Balmer line, the mean values of the two associated continuum regions were used to calculate a linear normalization line. The measured flux values were divided by this line, leading to the normalized line profile. Problems using the normalization occurred at spectra with low flux close to zero. Noise occasionally led to nonphysical values below zero, which jeopardized the normalization efforts. Such spectra were not further considered in the analysis.

\begin{table}
\caption{Line and continuum regions for Balmer lines.}
\centering
\small
\begin{tabular}{lccc} \hline \hline
Line      & Continuum 1 ($\text{\AA}$) & Line region  ($\text{\AA}$) & Continuum 2 ($\text{\AA}$) \\ \hline
$H\alpha$ & 6500 - 6550         & 6557.61 - 6571.61    & 6575 - 6625         \\
$H\beta$  & 4810 - 4850         & 4855.72 - 4870.00    & 4880 - 4900         \\
$H\gamma$ & 4270 - 4320         & 4331.69 - 4350.00    & 4360 - 4410         \\ \hline
\end{tabular}\tablefoot{All values adopted from \citet{hilton_2010AJ....140.1402H}}\label{tab:normalizatio_ranges}
\end{table}

%%%%%%%%%%%%%%%%%%%%%%METHODS%%%%%%%%%%%%%%%%%%%
\section{Methods}\label{sec:methods}

\subsection{Flare detection methods}\label{subsec:flaredetmethod}
To search for stellar flares in the spectra, we used methods based on identifying variations in the Balmer lines. The following subsections describe our approaches in more detail.

\subsubsection{Amplitude variability search: High Threshold Method (HTM) and Low Threshold Method (LTM)}\label{subsubsec:peakvar_search}

Using the normalized coadded spectra and the normalized single spectra, the $H\alpha$ and the $H\beta$ Balmer lines were fit by a simple Gaussian curve,
\begin{equation}\label{eq:gauss}
   I = 1 + I_{0} \times exp\left(\frac{(\lambda - \lambda_{0})}{\sigma} \right)^{2}
.\end{equation}
Here, $I_{0}$ denotes the line amplitude, $\lambda_{0}$ the wavelength of the line center, and $\sigma$ the width of the Gaussian profile. These three parameters were fit, and reasonable initial start values ($I_{0} = 0.0$, $\lambda_{0} = $ Balmer line core value, $\sigma = 4\: \mathrm{\text{\AA}}$) were chosen. The initial amplitude value was set to zero to account for absorption lines (negative $I_{0}$) and emission lines (positive $I_{0}$). All values around $ \pm 10\:\mathrm{\text{\AA}}$ of the core wavelength of each Balmer line were used in the fit.

Activity in $H\alpha$ was defined as a significant change of the fitted Gaussian line amplitude $I_{0}$ from one time instant to another. The amplitude change was required to exceed the error on the amplitude $I_{0}$ as derived from the fits. This amplitude error was determined by Gaussian error propagation and includes the amplitude errors of both Gaussians that are compared to each other, as each fit gives an error value. 
Two thresholds were used: amplitude changes greater than twice (high-threshold method, henceforth abbreviated as HTM) or 1.5 times (low-threshold method, henceforth abbreviated as LTM) of the amplitude error. To rule out nonphysical line fits, the resulting shift of the line core in the x-axis was required to be less than $2\:\mathrm{\text{\AA}}$ from one time instant to another.

Because of the low S/N, spectral resolution, and the possibility of outliers in the data, a significant change of the $H\alpha$ line alone is not sufficient to declare the spectrum as flaring. The same algorithm was therefore used on $H\beta$ as well. Because the noise around $H\beta$ is higher than around $H\alpha$, this second condition further decreases the number of eligible events compared to using $H\alpha$ alone. A star was defined to be in flaring state when the fitted $H\alpha$ and $H\beta$ lines both show amplitude variations above the thresholds.

The reasoning behind this method is that the cores of Balmer lines follow Gaussian curves arising from Doppler broadening. The fit is robust against outliers (e.g., cosmic rays) and therefore represents the overall line shape more faithfully, while trends in the line wings and outside of the Balmer lines are ignored. Especially late-type M-stars show strong contamination of molecular bands near $H\alpha$, for example. In addition to that, it is unlikely that both $H\alpha$ and $H\beta$ simultaneously change significantly above the amplitude error without flaring activity.

\begin{figure*}
 \centering
 \includegraphics[width=17cm]{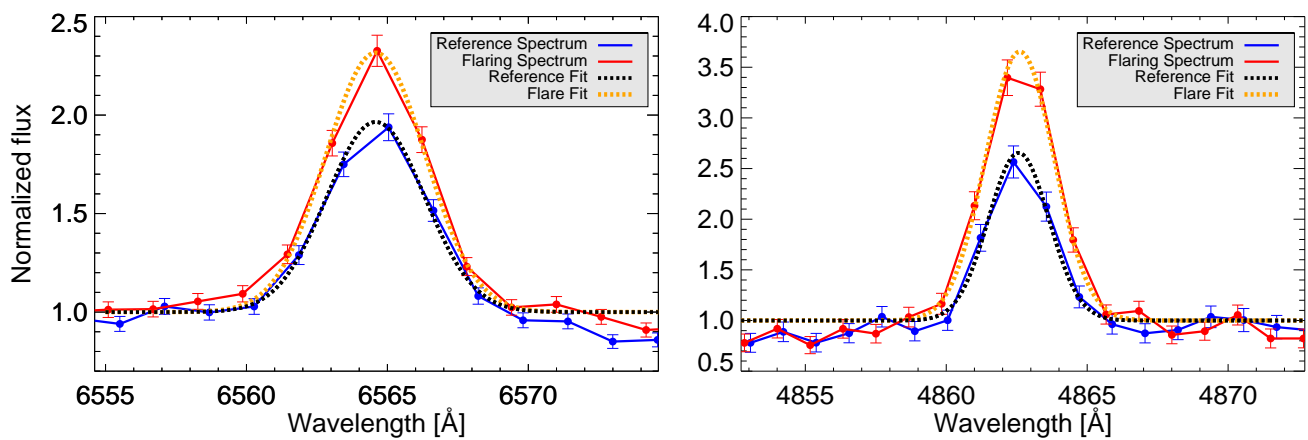}
 \caption{Example of a flaring $H\alpha$ line (left panel) and flaring $H\beta$ line (right panel) detected on a M5e star. Two single spectra are shown, one as a reference and one as an example of a flare. The data points with error bars together with the Gaussian fits are shown. \label{fig:example_gaussian_fit}}
 \end{figure*}

On the downside, the line-core fitting algorithm overestimates its fit significance for spectra with very low S/N by ignoring the noise around the Balmer lines. This made the visual post-inspection of each candidate event indispensable (see Section \ref{subsubsec:manualflaresearch}). Large outliers and faulty or missing data points could also lead to erroneous fits. Because of the spectral resolution of the SDSS spectra, the Balmer line cores (especially for M stars) usually consist of fewer than ten data points. Cosmic rays usually affect between 2 and 4 pixels of the camera on the spectrograph. Depending on the position and size of the cosmic ray hit, false data points may have similarities to emission lines. 
An example of a successful flare detection with this method is given in Fig. \ref{fig:example_gaussian_fit}. 

We applied this search strategy, and the HTM resulted in 538 flare candidates on M stars (183, 90, 73, and 192 flares for high, medium, low, and lowest S/N, respectively) and 47 candidates on K stars (19, 17, 6, and 5 flares for high, medium, low, and lowest S/N, respectively). The LTM detected an additional 420 M star flare candidates (9, 10, 35, and 366 flares for high, medium, low, and lowest S/N, respectively) and 132 on K  stars (18, 22, 15, and 77 flares for high, medium, low, and lowest S/N, respectively). After manual inspection, all of the 54 detections on G stars and 169 detections on F stars were considered to be either nonflaring  or had an incorrect spectral classification. The main reasons for this misidentification are either high noise, observational or data reduction errors that affect the whole spectrum, cosmic ray hits in both Balmer lines, the star being a close binary affecting the Balmer lines, or the star being a different type (e.g., RR Lyrae). 

\subsubsection{$H\alpha$ emission line method (ELM) }\label{subsubsec:haemission}
High noise and variations in the wings could lead to missing flares in the line-core fitting search algorithm. Because stars showing Balmer lines in emission are magnetically active and because the probability of detecting flares is therefore much higher, a search for stars showing the $H\alpha$ line in emission was started. We again used the Gaussian fitting algorithm (Eq. \ref{eq:gauss}), but this time, all stars were identified with an emission line amplitude $I_{0}$ higher than 1.5 at least. The resulting list of stars were then manually searched for changes  exceeding the typical errors from one time instant to another in both $H\alpha$ and $H\beta$. 

The $H\alpha$ emission line search algorithm for M stars found 177 in the high, 146 in the medium, 203 in the low, and 5049 stars in the lowest S/N bin. These numbers include several flaring stars that were previously detected with the HTM and LTM algorithms. For K stars, the search yielded 167 candidates (147 from the lowest S/N bin alone), but almost all of these detections are due to erroneous data values. There were no reasonable findings of strong Balmer emission in G and F stars except for detections arising from erroneous data. Detections in the lowest S/N bin were not further considered because the high number of false emission line detections in this list rendered a manual inspection unfeasible. After manual inspection, this method yielded six additional flares that were not found with the other methods.

\subsubsection{Visual inspection of the flare candidate spectra}\label{subsubsec:manualflaresearch}
After we obtained the numerous detections by the two algorithms, each identified candidate was verified manually to determine whether the flare was valid considering the following aspects: first we sorted out spectra with obvious data problems (e.g. incorrect flux levels due to data reduction errors or severe cosmic-ray hits). We also checked that the change in the line amplitude is significantly above the surrounding noise in both $H\alpha$ and $H\beta$. Considering the relative flare line amplitude and the error bars of the individual data points, we did not consider any amplitude changes that were too small (similar height as the errors or smaller). In addition to that, we checked for visible activity in other Balmer lines (especially $H\gamma$). Stars with no visible activity (enhanced data points in the same observation) in other lines despite having sufficiently good S/N were sorted out. Data point flags for each $H\alpha$, $H\beta$, and $H\gamma$ line were controlled. While flares are usually automatically flagged by the SDSS, several other aspects also result in flagged data points. Individual flags were controlled and spectra with many flags were noted. Finally, we cross-checked with flares identified in \citet{hilton_2010AJ....140.1402H} to determine whether a flare was mistakenly rejected based on one of these aspects. These events were then reevaluated (henceforth denoted Reeval.), which happened in four cases.

Based on these aspects, each spectrum was categorized into one of three groups "Yes", "Probable" or "No" to indicate whether the spectrum should be considered or discarded from the flare list. The stars in the category "Yes" were considered to show distinct flares. The spectra in the "Probable" category show clear activity in (at least) $H\alpha$ and $H\beta$, but more caution in handling them is appropriate. They either show no visible activity in $H\gamma$ (due to high noise) or several or all of the data points in one emission line were flagged, indicating a problem with this part of the spectrum. In these cases, all Balmer lines in the blue spectrum as well as the Ca II H and K lines were checked. Only events that clearly resembled usual flaring behavior were kept \citep[e.g., simultaneous flux increase in all Balmer lines instead of flux changes at different observation times; Ca II K line peaking later than the Balmer lines; see also ][]{kowalski_time_resolved_m_flares_2013ApJS..207...15K}. Four "Probable" flares were also reported in \citet{hilton_2010AJ....140.1402H}. Spectra with unusual properties (e.g. shifted Balmer lines) were checked regarding their spectral type classification. Eight stars (mis-) classified as M dwarfs by SDSS were identified as either cataclysmic variables (CV) or close/eclipsing white dwarf - M star binaries and were removed from the flare event list. CVs can be identified by the shift in their emission lines within a short period of time and the strong periodic changes in the strength of the lines.

Based on all these aspects, the final list of 281 flare candidates was created. This is the basis for further in-depth analysis.

\subsection{CME detection methods}\label{subsubsec:cme_search}
To search for signatures of stellar CMEs, we used methods based on the idea of identifying time variations in the Balmer line wings. This indicates motions of CME material with a detectable line-of-sight velocity component. The following subsections describe our approaches in more detail.

\subsubsection{CME-train search}\label{subsubsec:cme-train}
To detect enhanced emission or absorption in the blue or red wing of the emission or absorption line under study, two wavelength windows were defined covering the far red and blue wings of $H\alpha$ and $H\beta$ (corresponding to Doppler-velocity ranges of $\sim 350 - 1500\: \mathrm{km\:s^{-1}}$). A search was conducted to determine whether several successive data points (a "train") changed significantly from one time instant to another. All trains that exceeded a predefined length were reported. Train sizes of $5\: \mathrm{\text{\AA}}$ and $3\: \mathrm{\text{\AA}}$ were tested (equivalent to velocity ranges of $\sim 230 \: \mathrm{km \:s^{-1}}$ and $\sim 140 \: \mathrm{km \:s^{-1}}$ in $H\alpha$ and $\sim 310 \: \mathrm{km \:s^{-1}}$ and $\sim 180 \: \mathrm{km \:s^{-1}}$ in $H\beta$). The finding of one train in a spectrum triggered a detection that was subsequently cross-checked manually.

The idea behind this method was to find wide CME-related emissions like those found by \citet{houdebine_cme_1990A&A...238..249H}. The algorithm was successfully tested on an CME-related blue wing feature in the data from V374 Peg presented in \citet{vida_v374peg_cme_activity_2016A&A...590A..11V}. We note that this algorithm has two weaknesses: slow or small changes over several consecutive spectra are not recognized, and one single data point in the whole enhanced train with larger error or with an erroneous value may prevent a detection. 

This algorithm yielded 375 detections, which only consisted of either unusually large and wide cosmic-ray hits or spectra showing incorrect flux levels at one observation time due to faulty data reduction. A short discussion of the reasons for nondetections of CMEs is given in Sect. \ref{subdsubsec:signif:cme_discussion}

\subsubsection{Manual search for flare events and emission lines}\label{subsec_manual_cme_search}

\begin{figure*}
 \centering
 \includegraphics[width=17cm]{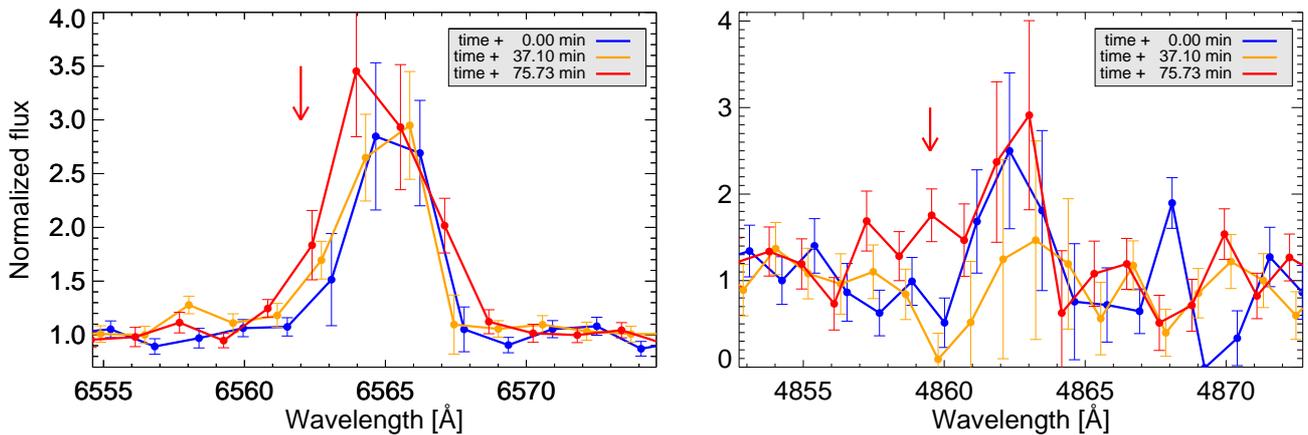}
 \caption{Example of possible blue-wing line asymmetries (indicated by arrows) and a possible flare in $H\alpha$ (left panel) and $H\beta$ (right panel) on an M4 star. This object has a low S/N and shows too little activity to be categorized as a CME or flare candidate. \label{fig:example_asymmetry}}
 \end{figure*}

All spectra found by the flare detection algorithms were inspected manually to search for Balmer line asymmetries or line wing enhancements or absorptions as an indicator of a CME. Every spectrum showing an asymmetry in either $H\alpha$ or $H\beta$ was noted (even when it was one data point). Spectra with clear asymmetries in both lines were recognized as possible CME candidates and analyzed in more detail. In these cases, the $H\gamma$ line and additional Balmer lines were included in the analysis, while all data point flags and error bars were controlled. Figure \ref{fig:example_asymmetry} shows an example of a low S/N star where an asymmetry was suspected. We found six possible CME candidates with this method (Sect. \ref{cme_analysis}). 

%%%%%%%%%%%%%%%%%%%%ANALYSIS-RESULTS%%%%%%%%%%%%%%%%%%%
\section{Analysis and results}\label{sec:analysis}

\subsection{Flare-search results}
Using these methods, we detected no flares for dwarf stars of types G or F, 272 events were found for M dwarfs, 6 on K dwarfs, and 3 on M-type stars that are known T Tauri stars. These 3 stars are labeled primarily by their luminosity class (T Tauri) from now on to differentiate them from the M dwarfs. This gives us a final list of 281 flare candidates.

Table \ref{tab:flares_bins} shows all flares per spectral subtype listed for each S/N bin. One of the K stars is most likely a giant or subgiant star; this detail as well as the ambiguity of the other K-type stars is discussed in Appendix \ref{sec:appendix_special_cases_flares}. Table \ref{tab:flares_methods} lists the final number of flares that were found by each method, showing that HTM proved to be the most successful flare-search method in this work.

\begin{table*}
\caption{Spectral type of flaring stars per S/N bin.}
\makebox[\textwidth][c]{
\begin{tabular}{l|cccccccccccccc|c} \hline \hline
\begin{tabular}[c]{@{}c@{}} S/N bin \end{tabular} & K3 & K5 & K7 & M0 & M1 & M2 & M3 & M4 & M5 & M6 & M7 & M8 & M9 & TT & Total \\ \hline
high & 1 & 2 & 3 & 1 & 0 & 4 & 6 & 35 & 29 & 19 & 5 & 0 & 0 & 2 & 107 \\
medium & 0 & 0 & 0 & 0 & 1 & 0 & 3 & 3 & 15 & 13 & 3 & 1 & 0 & 1 & 40 \\
low & 0 & 0 & 0 & 0 & 0 & 1 & 2 & 4 & 2 & 16 & 14 & 1 & 1 & 0 & 41 \\
lowest & 0 & 0 & 0 & 0 & 0 & 0 & 2 & 1 & 5 & 30 & 29 & 9 & 17 & 0 & 93 \\ \hline
Total & 1 & 2 & 3 & 1 & 1 & 5 & 13 & 43 & 51 & 78 & 51 & 11 & 18 & 3 & 281 \\ \hline
\end{tabular}}
\label{tab:flares_bins}
\end{table*}

\begin{table*}
\caption{Number of detected flaring stars per method.}
\makebox[\textwidth][c]{
\begin{tabular}{l|ccccc|c} \hline \hline
S/N bin & HTM & LTM & ELM & Reeval. & Other & Total \\ \hline
high    & 99                   & 5                    & 2             & 0                        & 1     & 107   \\
medium  & 38                   & 1                    & 0             & 1                        & 0     & 40    \\
low     & 32                   & 4                    & 3             & 2                        & 0     & 41    \\
lowest   & 67                   & 24                   & 1             & 1                        & 0     & 93    \\ \hline
Total   & 236                  & 34                   & 6             & 4                        & 1     & 281  \\ \hline
\end{tabular}}
\tablefoot{HTM and LTM: High- and low-threshold method, ELM: emission line method, and Reeval: reevaluated spectra. The one flare in the "Other" category was found on a K star by inspecting $H\alpha$ activity alone.}
\label{tab:flares_methods}
\end{table*}

\subsection{Flare parameters}\label{subsec:flare_parameters}

Efforts were made to constrain flare parameters, in particular the flare occurrence, flare energies, flare luminosities, and flare evolution for each identified flare. We briefly describe the preceding analysis and object parameters necessary to yield these results. Regarding the stellar distances, 259 ($92.2\%$) of the flaring stars have associated GAIA DR2 parallaxes (see Sect. \ref{subsec:gaiadr2}). Stars with missing parallaxes were checked for other sources within VizieR, and if possible, a value for the distance or the parallax was added. The spectral classification cross check (see Sect. \ref{subsec:spectral}) served as an indicator to determine whether the SDSS spectral classification was valid. We found 253 of the 281 flaring stars within the \citet{spectral_class_2014yCat....1.2023S} catalog. Missing stars were again checked for other sources. If not stated otherwise, the spectral types from the literature are used in the following subsections. 

It is necessary to define a reference spectrum for each star for comparison with the flaring spectra and to gather the flare parameters. A "quiet flag" was introduced to mark reference spectra that were in a nonflaring state. Spectra were considered to be quiet when at least two of them at a completely different observation time than the flaring spectra show negligible differences. Spectra were considered to be "probably quiet" when they appeared before or after a consecutive flaring series without any further flux changes. Of the 281 events, 59 show quiet spectra and 64 show probably quiet spectra.

\subsubsection{Flaring occurrence, positions in the color - magnitude diagram, and sample purity}\label{subsec:flaring_occurrence,cmd,smaple_purity}
To estimate the flare occurrence rates for each stellar subtype, the flaring times of all events were summed for each subtype and divided by the overall exposure times of all available stars of the corresponding subtype. The flaring percentage of the later M-type stars is highly dependent on whether a possible flare was detected despite the poor S/N. Not enough flares were found on K-type stars for meaningful statistics.

\begin{figure}
 \centering
 \resizebox{\hsize}{!}{\includegraphics{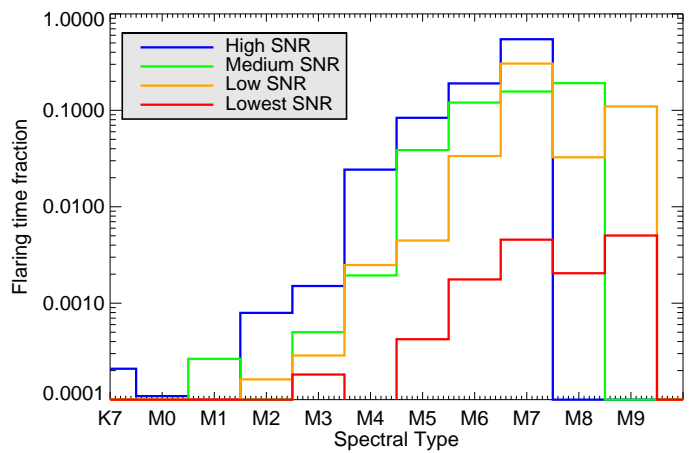}}
 \caption{Flaring occurrence per stellar subtype in percent (fraction of observed flaring time compared to overall exposure time in the dataset for each spectral subtype). The S/N bins are color-coded.
\label{fig:flaring_percentage}}
 \end{figure}

Figure \ref{fig:flaring_percentage} shows the results of the flaring occurrence separately for the spectral subtypes and S/N classes. The high and medium S/N bins are more reliable than the other bins. No flare on an M8 or M9 star was detected in the high S/N class, which is again a consequence of their low flux. Overall, the flaring occurrence appears to be higher for later type stars.\\
It is important to note that the overall exposure times per subtype were based on the original SDSS classification, while the flaring stars were classified using several sources. The large flaring fraction reaches high values (up to $55\:\%$ for type M7) because very few late-type stars lie in these S/N bins (only $\text{about ten}$ stars of type M7 stars are in "High"). Here, the mix of different spectral classifications may compromise the absolute values, but the overall trend is valid.

Using the data given by GAIA DR2, we can plot the position of the flaring stars in the color - magnitude diagram (CMD). The basic idea behind this is to test whether the stars in question are in fact late-type dwarf stars, resulting in a high sample purity regarding the luminosity class.

\begin{figure}
 \centering
 \resizebox{\hsize}{!}{\includegraphics{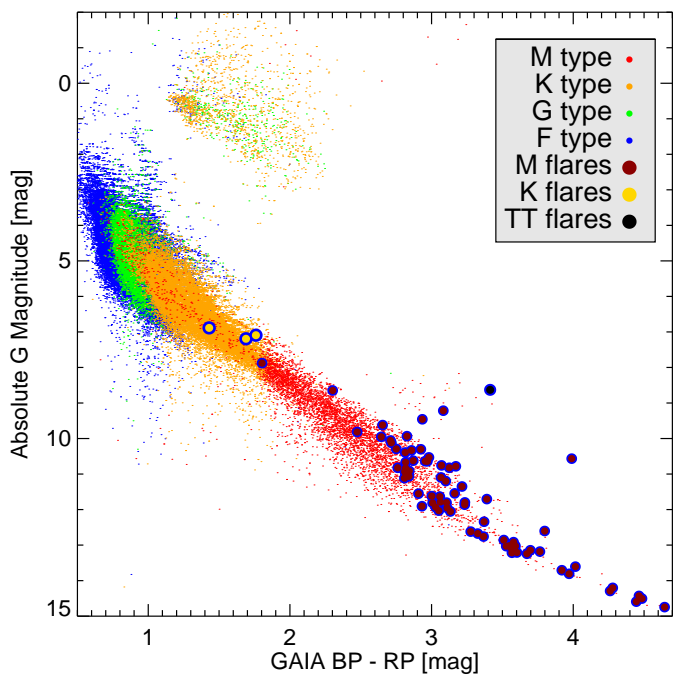}}
 \caption{CMD of the SDSS sample with errors $< 1\:$ mag in both axes for the original sample. The flaring subsample has errors of $< 2\:$ mag in absolute magnitude and $< 1\:$ mag in the color axis. The main constraining factor is the availability of error values for the color. \label{fig:hrd_gaia_stars:flares}}
 \end{figure}

Figure \ref{fig:hrd_gaia_stars:flares} shows the flaring stars in the CMD based on their absolute G magnitude derived from GAIA DR2 and their GAIA BP - RP color. The respective errors for the BP - RP color as well as the absolute magnitude were restricted to be smaller than 1 mag. This restriction reduces the whole SDSS sample to approximately 121000 stars ($\sim 19\:\%$ of all available stars) and the number of flaring stars to 76 ($30 \%$ of all flaring stars) in the plot. This reducted number of stars in this plot is mostly due to large errors or missing values in the BP - RP value from GAIA. We find that most of the displayed stars lie well within the main sequence. Some stars are above the M star main-sequence region, which is an indicator for pre-main-sequence stars (T Tauri stars), one of which was confirmed to be one. None of the flaring stars presented in this plot shows signs of being an evolved star on the RGB.

\begin{figure*}
 \centering
 \includegraphics[width=17cm]{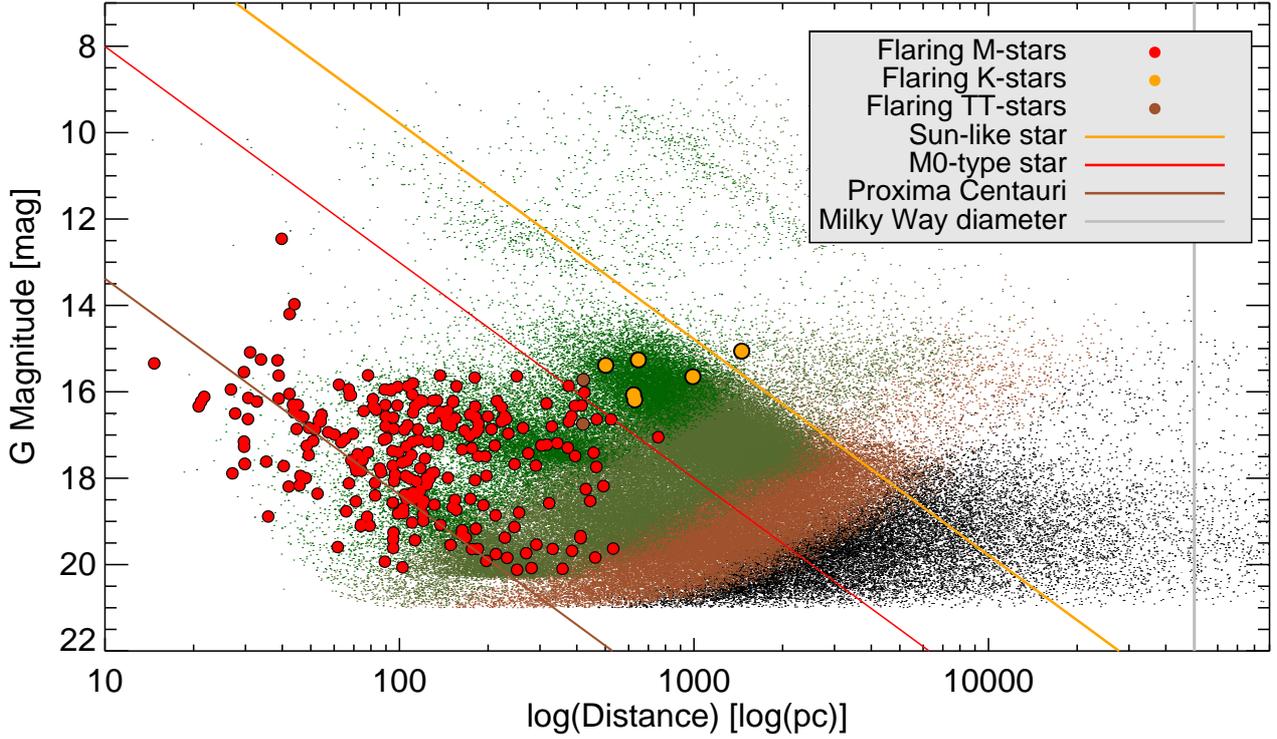}
 \caption{M- and K-type stars (using the SDSS spectral type classification) with their respective distances and G magnitudes from GAIA DR2. Flaring stars are shown within the plot. Example for a dwarf star (Proxima Centauri), an M0-type star (GJ 270), and a Sun-like star are given. The diameter of the Milky Way is given as an upper constraint on the distance. Errors in the distance are color-coded. Green shows error smaller than $5 \%$. Olive shows an error between $5$ and $20 \%$. Brown shows an error between $20$ and $50 \%$. Black shows an error higher than $50 \%$. \label{fig:sample_purity}}
 \end{figure*}

Figure \ref{fig:sample_purity} shows all M and K stars in the dataset with their GAIA G magnitudes over their distances. All flaring stars are displayed as well, and all flaring M stars lie well within 1000 parsec. The G magnitude of a solar-like star was plotted as a function of the distance to give a distinction line between stars with subsolar luminosity (such as M and K dwarfs) and stars with higher than solar luminosity (like red giants). One instance of a flaring K star above this line is a clear indication that this star is either a subgiant or a giant star (other possibilities include T Tauri stellar type). The function of a well-known M5.5 dwarf (Proxima Centauri) is also displayed. The function of an M0 type star (GJ 270) is displayed to approximately show the threshold between K and M stars. The error of the distance is color coded as described in the figure caption. The diameter of the Milky Way is given as an upper limit for unrealistic distance values. The plot shows that we obtained the object information out of two different datasets (SDSS and GAIA DR2). The lower threshold in magnitude is most obvious because GAIA DR2 restricts its measurements of stars brighter than 21 mag in G \citep{2018gaiadr2}.  Figs. \ref{fig:hrd_gaia_stars:flares} and \ref{fig:sample_purity} show the observed stellar magnitudes without a correction for extinction, which would affect the true magnitude values of the most distant stars. Figure \ref{fig:sample_purity} further indicates that the displayed sample of flaring M stars only consists of M dwarfs. It also works as a tool for detecting giant stars in the flaring K star subsample. For a short discussion of the sample purity, see Sect. \ref{subsubsec:discussion:sample_purity}.

\subsubsection{Flare energies, flare luminosities, and detection threshold}\label{subsec:energy_calc}
We calculated the flare energy $E_{\rm flare}$ and peak luminosity $L_{\rm flare}$ in $H\alpha$. Other chromospheric lines were not considered because their S/N is lower than in $H\alpha$.

As the data values in the spectra are given as spectral flux density $I_{\rm{\lambda}}$ in units of [$10^{-17}\mathrm{\:erg \: s^{-1} cm^{-2} \text{\AA}^{-1}}$], we calculated the fluxes $F$ ([$\mathrm{erg \: s^{-1} cm^{-2}}$]) for each spectrum (abbreviated by "spec") by
\begin{equation}\label{equ:radiance}
   F_{\rm spec} =\int_{\lambda_{1}}^{\lambda_{2}} I_{\rm{\lambda}}(\text{\AA}) d\text{\AA}
,\end{equation}{}
where $\lambda_{1}$ and $\lambda_{2}$ determine the wavelength range for the integration. The window around $H\alpha$ is $6555 - 6575\: \mathrm{\text{\AA}}$. Prior to the integration, linear interpolation was used. This was necessary as each spectrum usually has a different wavelength axis, which often results in a different number of data points in the integration range. 
The excess flaring flux was obtained by subtracting the reference, $ F =  F_{\rm flare} - F_{\rm ref}$. 
The reference (ref) spectrum was either a spectrum declared as quiet, the coadded spectrum, or the spectrum showing the least amount of activity. Spectra showing cosmic rays in the integration range as well as flares with high noise in the reference were excluded. The total flare flux (sum of all flaring fluxes of an event) as well as the peak flaring flux were evaluated. The total radiant energy per area $H$ ([$\mathrm{erg \: cm^{-2}}$]) was obtained by multiplying the flaring excess flux of each spectrum with the exposure time and summing,
\begin{equation}\label{equ:rad_energy}
   H = \sum_{\rm i}^{\rm N} F_{\rm{i}} * t_{\rm exp, \: i}
.\end{equation}{}
Here, $i$ denotes each single spectrum, and $N$ is the total number of consecutive flaring spectra.
The total flare energy (in units of [erg]) is calculated by
\begin{equation}\label{equ:energy}
 E_{\rm flare} =  H * 4\pi d^{2}
,\end{equation}
and the instantaneous flare luminosity (in units of [$\mathrm{erg\:s^{-1}}$]) is given as 
\begin{equation} \label{equ:power}
   L_{\rm flare} =  F * 4\pi d^{2}
,\end{equation}
with $d$ being the distance to the star. $E_{\rm{flare}}$ (as well as H) gives a total value for the flare energy, while $L_{\rm {flare}}$ gives a luminosity value for a single flaring spectrum. 

The integration errors for the energy and luminosity were determined as follows:
\begin{equation}\label{equ:errors_integrated}
    \Delta F_{\rm spec} = \Delta \lambda * \sqrt{\sum_{\rm j}^{\rm M} (error_{\rm (j)})^{2}}
,\end{equation}{}
with $\Delta \lambda$ being the mean wavelength spacing of the data points, $M$ the number of all data points in the range (without interpolation), and $error_{\rm{(j)}}$ the error bar associated with data point j. Using Gaussian error propagation,
\begin{equation}\label{equ:errors-power_integrated}
    \Delta F = \sqrt{\Delta F_{\rm ref}^{2} + \Delta F_{\rm flare}^{2}}
\end{equation}{}
and
\begin{equation}\label{equ:errors-energy_integrated}
   \Delta H = \sqrt{ \sum_{\rm i}^{\rm N} \left(\Delta F_{\rm i} *t_{\rm exp, \: i}\right)^{2}} ,
\end{equation}{}
 the final result of the error in flare energy was determined by
\begin{equation}\label{equ:error_energy}
    \Delta E_{\rm flare} = \left|\Delta H \: 4\pi d^{2}\right| + \left|\Delta d \: \: H * \: 8\pi d\right|
,\end{equation}{}
and for the luminosity by
\begin{equation}\label{equ:error_power}
    \Delta L_{\rm flare} = \left|\Delta F \: 4\pi d^{2}\right| + \left|\Delta d \: \: F \: 8\pi d\right|
,\end{equation}{}
with $\Delta d$ being the associated error determined by the GAIA DR2 parallax. If no distance error value was available, the second term in Eqs. \ref{equ:error_energy} and \ref{equ:error_power} was neglected.

\begin{figure}
 \centering
 \resizebox{\hsize}{!}{\includegraphics[width=1.0\columnwidth]{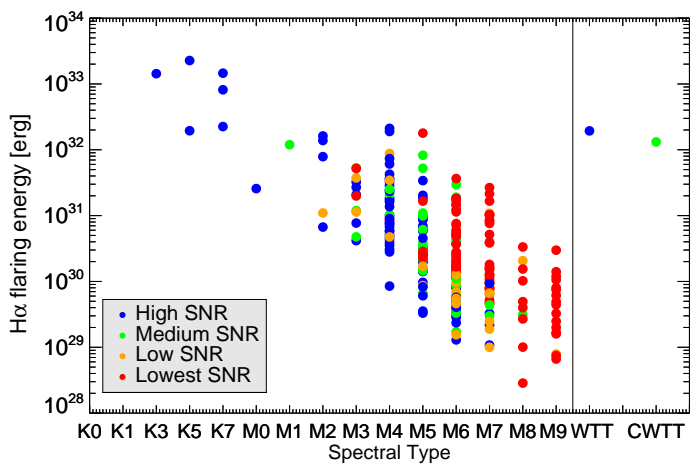}}
 \caption{Flare energy as a function of stellar subtype. T Tauri stars are displayed separately. The S/N bin for each star is given.
  \label{fig:stellar_flare:_energy_subtype_snr}}
\end{figure}

Figure \ref{fig:stellar_flare:_energy_subtype_snr} shows the result of the $H\alpha$ flare energy over the spectral subtype. The flare energy arising from excess flux in $H\alpha$ ranges over five orders of magnitude from $\sim3\times 10^{28}\: \mathrm{erg}$ up to $2\times 10^{33}\: \mathrm{erg}$. The plot shows a clear trend in the flare energy, with late-type stars showing less energetic flares than earlier stars. S/N bins are color-coded for each flare. The highest calculated flare energies are observed on K stars and earlier M stars. This is probably a selection effect: because K stars are brighter, they need higher energy releases for flares to be detected. Another explanation for this trend is the simple correlation with distance: there are more late-M stars in the solar neighborhood, making their flares (even at small energies) easier to detect, while there are no late-M dwarf stars in the dataset with large distances to the Solar System. Their flux is not high enough for spectroscopic observation in the SDSS. K stars, on the other hand, show higher absolute magnitudes and are easier to observe at larger distances.

\begin{figure}
 \centering
 \resizebox{\hsize}{!}{\includegraphics{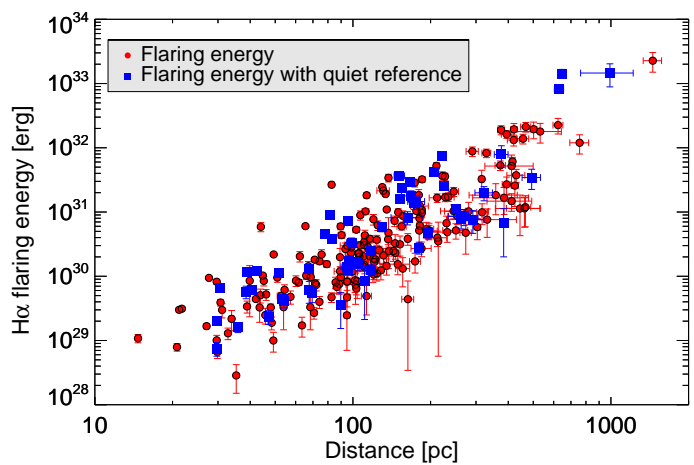}}
 \caption{$H\alpha$ flare energy as function of distance from the GAIA DR2 catalog. Stars with quiet reference spectra are marked in blue. Stars where the reference spectra is not unambiguously in a quiet state are marked in red. \label{fig:stellar_flare:_energy_distance}}
 \end{figure}

Figure \ref{fig:stellar_flare:_energy_distance} shows the result of the $H\alpha$ flare energy over the distance. Stars that have quiet reference spectra are marked in blue. The energy values of these flares are more reliable because the reference is not part of the flare, which would compromise the calculation. It was not clear whether the reference spectrum was quiet or still in a flaring state for all flares marked in red. The two groups clearly agree, indicating that the obtained values marked in red are also fairly reliable.

We again found the same range for the flare energy arising from excess flux in $H\alpha$ with the typical spread of about one order of magnitude in energy. 
The more distant a star, the more energetic the flare has to be to be detected with our methods. This is also another validation of the significance of the flares that we found. This plot mostly depends on the reliability of the GAIA parallax information. Calculated error bars (using Eq. \ref{equ:error_energy} for the energy and GAIA parallax errors for the distance) are displayed. The apparent upper limit of energetic flares (showing that there appear to be no energetic flares at a close distance) of nearby stars may be a result of a selection effect. This is discussed in Sect. \ref{subdsubsec:notes_energy}.

\begin{figure}
 \centering
 \resizebox{\hsize}{!}{\includegraphics{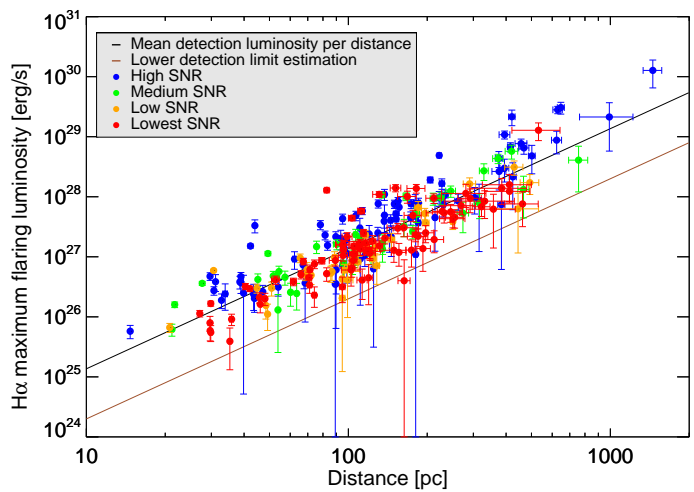}}
 \caption{Flare peak luminosity as a function of stellar distance. S/N bin of each star and the respective error bars are color-coded. The mean detection luminosity at a given distance as well as an estimate of the lower detection limit is displayed by the black and brown lines.    \label{fig:flare_peak_distance}}
 \end{figure}

To answer the question which minimum flare threshold is still detectable in this dataset, we considered the relation between the maximum calculated $H\alpha$ flare luminosity (in units of [erg/s]) and the distance of each flare star. A simple quadratic equation was used,
\begin{equation}\label{equ:quadratic_fit}
    L_{\rm fit}(d) = 10^{c} \times d^{2}
.\end{equation}{}
Here, $L_{\rm fit}$ denotes the fitted function of flare luminosity as a function of the distance d in parsec, and $c$ is a free parameter determining the detection threshold. The coefficient $10^{c}$ is given in units of [$\mathrm{erg/s/pc^{2}}$]. For the mean flare luminosity as a function of distance, a value of $c= 23.1344$ was found. The lower threshold was estimated with a value of $c \sim 22.3$. This means that a star at 100 pc distance needs a measured $H\alpha$ flare luminosity of at least $10^{22.3}\times 100^{2} = 10^{26.3}\: \mathrm{ erg/s}$ to be detected in this dataset (see Fig. \ref{fig:flare_peak_distance}).

\subsubsection{Flare evolution}\label{subsec:flare_evolution}
The $H\alpha$, $H\beta$, and $H\gamma$ Balmer lines were manually inspected. To estimate the evolution of the flare, every event was categorized as follows: "Rise" (Balmer lines grow over consecutive time series), "Decay" (Balmer lines decay over consecutive time series), "Full" (entire flare evolution including peak), "Activity at different time" (a single flaring spectrum at a separated observation date, preventing an evolution estimate), and "Other" (unable to classify the evolution of the flare). Irregular cases, that is, classes in which the flare decreases and rises again immediately afterward, fall into the "Other" category. At each instant, the whole available spectrum (especially other Balmer lines) was checked to gain further information.

The results of the manual flare evolution identification is summarized in Table \ref{tab:flare_evolution}. These results were cross-checked with the overlapping flares from \citet{hilton_2010AJ....140.1402H}, giving the same results. More flares are declared to be in a rising phase than in decay. This finding is discussed in more detail in Sect. \ref{subdsubsec:notes_flare_evol}.

\begin{table}
\caption{Flare evolution estimates of all 281 flares.}
\small
\begin{tabular}{c|ccccc|c} \hline \hline
       & Rise & Decay & Full & Activity at diff. time & Other & Total \\ \hline
$\#$ & 82   & 67    & 55   & 67                         & 10    & 281   \\ \hline
\end{tabular}
\label{tab:flare_evolution}
\end{table}

\subsection{CME search and results}\label{subsec_cme_candidates}

\subsubsection{Overview of CME candidates}\label{sec:cme_candidates}
We present six possible CME candidates, all detected on M stars, that show enhancements in Balmer line wings (see Sect. \ref{cme_analysis}). They were found by noting wing asymmetries in all stars that were manually inspected (mostly flare candidate stars). 
Five candidates show a red wing, and one shows a blue-wing enhancement. We discuss the interpretation of red-wing enhancements in Sect. \ref{subdsubsec:cme_findings}. An overview of these candidates is given in Table \ref{tab:cme_candidates_list}, including general information on the objects.

\begin{table*}
\caption{List of the CME candidates.}
\makebox[\textwidth][c]{
\small
\begin{tabular}{lccccccl} \hline \hline
\multicolumn{1}{l}{No} & SDSS ID & RA & DEC & Type & Flare & \multicolumn{1}{l}{S/N Bin}& Lines with wing enhancement \\ \hline
1 & J052700.12+010136.8 & 81.75054 & 1.02691 & M3.2e*$^{[1]}$ known WTT$^{[2]}$ & No & high& $\boldsymbol{ H\alpha}$, $H\beta$, $\boldsymbol{H\gamma}$ \\
2 & J085556.66+174942.0 & 133.98612 & 17.82835 & M4e$^{[3]}$ & Yes & high&$\boldsymbol{H\alpha}$, $H\beta$, $H\gamma$ \\
3 & J110743.30+075624.9 & 166.93047 & 7.94027 & M4e$^{[3]}$ & No & high& $H\alpha$, $\boldsymbol{H\beta}$, $\boldsymbol{H\gamma}$, $H\delta$, $H\epsilon$, $H8$ \\
4 & J042139.64+264913.8 & 65.41520 & 26.82050 & M6e*$^{[4]}$ & Yes & medium&$\boldsymbol{H\alpha}$, $\boldsymbol{H\beta}$, $H\gamma$, $H\delta$ \\
5 & J053709.91-011050.3 & 84.29134 & -1.18067 &M5e*$^{[5]}$ known WTT$^{[6]}$ & Yes & high&$\boldsymbol{H\alpha}$,$\boldsymbol{H\beta}$,$\boldsymbol{H\gamma}$,$\boldsymbol{H\delta}$,$\boldsymbol{H\epsilon}$,$H8$,$Ca II H \& \boldsymbol{K}$ \\
6 &J052839.66-000322.5 & 82.16528 & -0.05627 & M4e$^{[3]}$ & Yes & high& $\boldsymbol{H\alpha}$, $\boldsymbol{H\beta}$, $H\gamma$, $\boldsymbol{H\delta}$ \\ \hline
\end{tabular} }
\tablefoot{Listed properties including SDSS ID, right ascension and declination, spectral subtype, if an associated flare is visible, S/N bin, and the lines showing the enhancement. Significant enhancements in line wings (above all error bars) are marked in bold font. Spectral type sources: $^{[1]}$ \citet{suarez_low_mass-Stars_ori_spectraltype_2017AJ....154...14S}, $^{[2]}$\citet{mcgehee_origroup_stars2006AJ....131.2959M}, $^{[3]}$\citet{west_m_dwarfs_sdss_spectral_types_distances_2011AJ....141...97W}, $^{[4]}$\citet{luhman_taurus_starforming_region_2017AJ....153...46L}, $^{[5]}$\citet{brieceno_orion_survey_variables_2019AJ....157...85B}, $^{[6]}$\citet{koenig_ysos_in_orion2015AJ....150..100K}. *The "e" for emission line star was added to these stars based on the visual inspection.}
\label{tab:cme_candidates_list}
\end{table*}

\begin{table*}
\caption{CME candidates with associated values.}
\makebox[\textwidth][c]{
\small
\begin{tabular}{ccccccclcl} \hline \hline
No & Enhanced&Distance&Mean $H\alpha$&Max. velocity& Bulk velocity & Est.mass $H\alpha$&Error &  Est.mass $H\gamma$&Error \\ 
& Wing & [pc] & S/N & [$\mathrm{km~s^{-1}}$] & [$\mathrm{km~s^{-1}}$] & [g] & \% & [g] &\%  \\ \hline
1 & red & 379.3&28 & 594& 494 & 3.0E+17 &81& X&  X \\
2 & red & 234.2&22 & 708& 275 & 6.4E+17 &22& X& X \\
3 & red & 70.5&28 & 667& 300 & 6.1E+16 &51& 1.3E+17&24 \\
4 & red & 166.3&17 & 703& 358& 8.5E+17 &11& X &X\\
5 & red &   -   &21  & 708& 340 & - & - & - & - \\
6& blue & 467.8&24 & -365& -240&  6.1E+18 &19& X& X \\ \hline
\end{tabular} }
\tablefoot{Listed properties include indication of the direction of the Doppler shift, the distance in parsec based on Gaia DR2, their maximum and bulk velocity, and the estimated mass (including rough error estimates in percent) using the flux power in $H\alpha$ and $H\gamma$. A cross marks cases where the mass estimation was not possible for specific reasons (see Sect. \ref{cme_analysis}).}
\label{tab:cmes_calculated_values}
\end{table*}

\subsubsection{Mass and velocity estimations}\label{subsec_mass_vel_cmes}
Under the assumption that the line wing enhancements we found are due to CME material having a velocity component along the line of sight, we can estimate the velocity of the outflowing material. The maximum speed was estimated as the point at which the enhancement merges with the continuum, while the bulk velocity was determined by fitting a Gaussian to the enhanced feature, determining its central wavelength. If a Gaussian fit to the enhanced line wing feature was not possible, the bulk velocity was estimated using the mean wavelength value of the enhancement.

To estimate the hydrogen mass, the following relation from \citet{houdebine_cme_1990A&A...238..249H} was used:
\begin{equation}\label{equ:houdebine:mass_cme}
    M_{\rm CME} \geq \frac{4\pi d^{2} F_{\rm em} \left( N_{\rm tot}/ N_{\rm j} \right) m_{\rm H} \eta_{\rm OD}}{h \nu_{\rm j-i}A_{\rm j-i}}
.\end{equation}{}
Here, \textit{d} denotes the distance of the star, $F_{\rm em}$ the integrated excess flux from the emission feature, $N_{\rm tot}/N_{\rm j}$ the ratio between the total number of hydrogen atoms to the hydrogen atoms being at excitation level \textit{j}, $m_{\rm H}$ the mass of the hydrogen atom, $\eta_{\rm OD}$ the opacity damping factor (giving a value to indicate how much flux escapes the plasma due to optical thickness), \textit{h} the Planck constant, $ \nu_{\rm j-i}$ the frequency difference between excitation level \textit{j} and \textit{i}, and $A_{\rm j-i}$ the Einstein coefficient for spontaneous decay from excitation level \text{j} to \text{i}.

While the CME candidates are most prevalent in the $H\alpha$ line (transition from j= 3 to i = 2), no value is available in the literature to estimate the ratio $N_{tot}/N_{3}$. However, the transition for $H\gamma$ (j = 5, i = 2) is estimated to be $N_{tot}/N_{5} = 2 \times 10^{9}$ by \citet{houdebine_cme_1990A&A...238..249H} using nonlocal thermal equilibirum (NLTE) modeling for the active M star AD Leo. $\eta_{OD}$ was set to 2, indicating that $50\: \%$ of the radiation escapes the plasma \citep{leitzinger_flare_cme_search_blanco_2014MNRAS.443..898L}, while the Einstein coefficient is $A_{5-2} = 2.53 \times 10^{6}$  \citep{einstein_coef_wiese2009accurate}.

As the SDSS spectra cover the whole visible spectrum, the $H\gamma$ line can be used, but poses difficulties because the S/N is low, especially for faint late-type stars. We can therefore approximate the real value of the CME flux in $H\gamma$ by scaling the flux from $H\alpha$ with the Balmer decrement BD ($F_{em,\gamma} = F_{em,\alpha} / BD$), with BD set to 3 \citep[adopted from solar and stellar flares,][]{butler_balmer_decrement_1988A&A...206L...1B}. When possible, we used both $H\alpha$ and $H\gamma$ measurements for CME mass estimations for each candidate.

For error estimations, the same method as in the flare flux error analysis was used (see Eq. \ref{equ:errors_integrated}). For all CME excess flux measurements, the flux as well as the errors of the coadded spectrum was used as reference. The coadded spectrum has smaller uncertainties because the S/N is higher.

Table \ref{tab:cmes_calculated_values} shows calculated values such as maximum velocity, bulk velocity, and mass estimations based on the CME assumptions. The derived maximum velocities range from 365 $\mathrm{km~s^{-1}}$ to 700 $\mathrm{km~s^{-1}}$ and the  bulk velocities from 240 $\mathrm{km~s^{-1}}$ to 500 $\mathrm{km~s^{-1}}$ (absolute values). The derived CME masses range from $6 \times 10^{16}\:$g to $6 \times 10^{18}\:$g.

\subsubsection{Detailed analysis of the six CME candidates showing enhanced Balmer line wings}\label{cme_analysis}

\begin{figure}
 \centering
 \resizebox{\hsize}{!}{\includegraphics{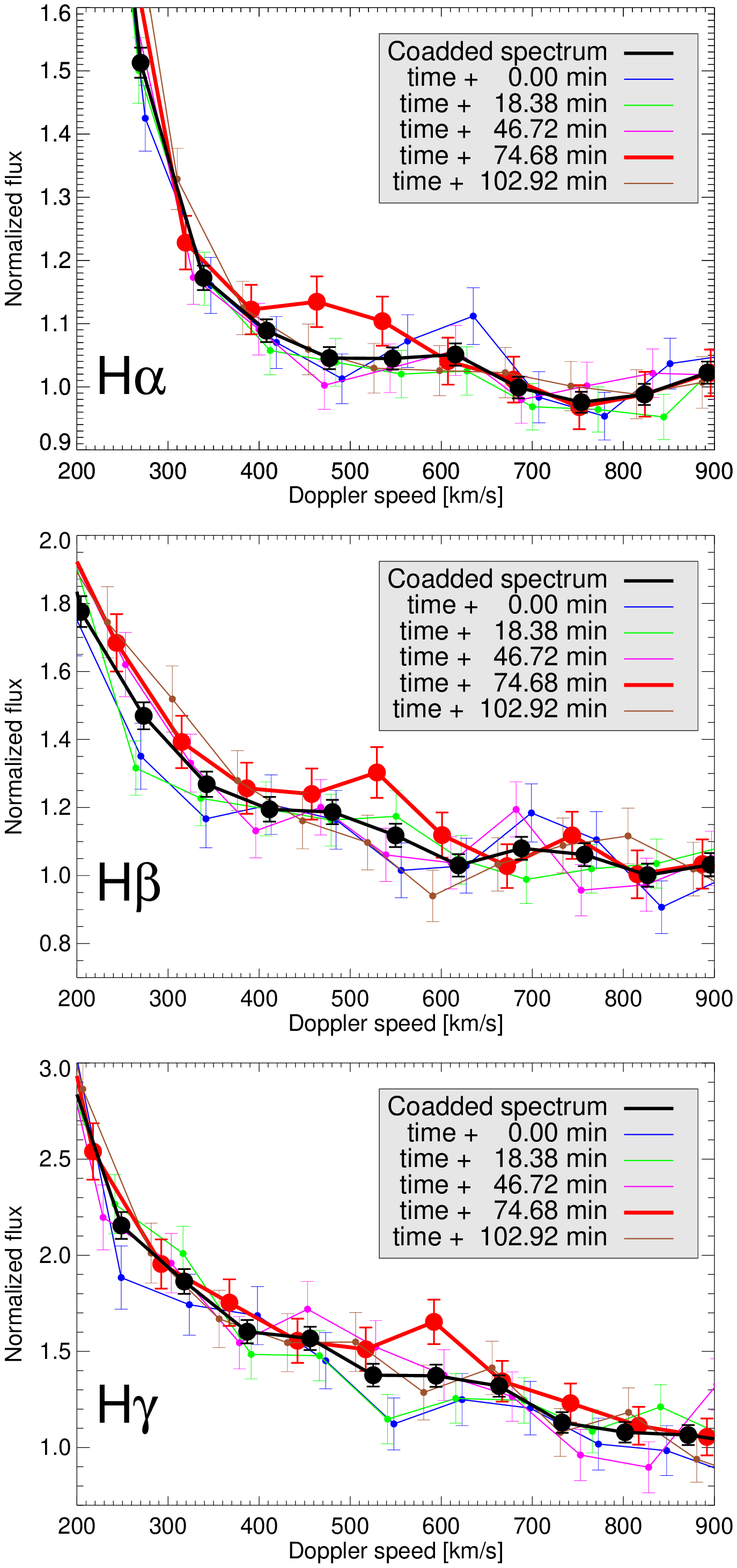}} \caption{Red wing of CME candidate 1 in $H\alpha$ (upper panel), $H\beta$ (middle panel), and $H\gamma$ (lower panel) for five time steps. The spectral line at the time step showing the enhancement is marked in red.}  \label{fig:cme_1}
 \end{figure}

\begin{figure}
 \centering
 \resizebox{\hsize}{!}{\includegraphics{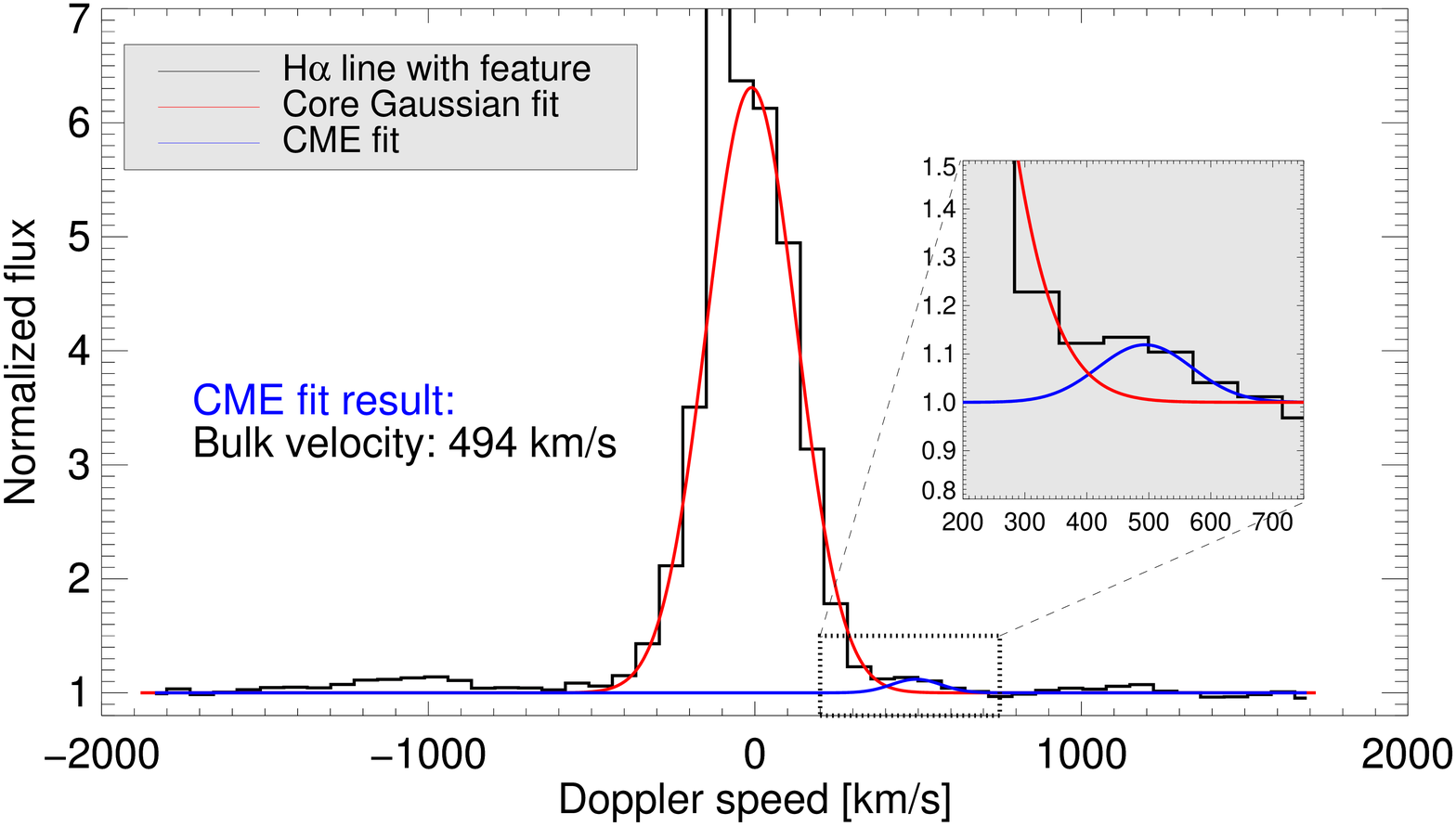}}
 \caption{$H\alpha$ spectral line of CME candidate 1. The line core (red line) as well as the possible CME feature (blue line) is fitted, with the bulk velocity resulting from the Gaussian CME fit annotated. A cosmic-ray hit affects the spectrum near the $H\alpha$ line peak. The inset shows a zoom into the spectral range with the enhanced wing. \label{fig:cme_1_fitted}}
 \end{figure}

\textbf{Candidate 1: SDSS J052700.12+010136.8 (Figs. \ref{fig:cme_1_fitted} and \ref{fig:cme_1}) }

The star with subtype M3.2 (SDSS classification M4) was found with our ELM algorithm. \citet{mcgehee_origroup_stars2006AJ....131.2959M} also reported it as a weak-line T-Tauri star (WTTS). The SDSS spectrum shows five consecutive spectra with exposure times of 15 minutes for spectra 1 and 5, and 25 minutes for spectra 2, 3, and 4. Figure \ref{fig:cme_1} shows the red wing of the $H\alpha$, $H\beta$, and $H\gamma$ lines of these spectra\footnote{We note that the enhancement in the spectrum at about $ -1000 \: \: \mathrm{km~s^{-1}}$ is due to a molecular band that is not considered in the linear normalization procedure and thus can be ignored. The same holds for Figs. \ref{fig:cme_2_fitted}, \ref{fig:cme_4_fitted}, \ref{fig:cme_5_fitted}, and \ref{fig:cme_6_fitted}. }. Spectrum 4 shows a significant increase in flux on two neighboring data points in the red wing of $H\alpha$ as well as one (not significantly enhanced) higher data point on $H\beta$ and one significantly enhanced data point in $H\gamma$. We recall that the wavelength axis is oversampled with respect to the spectral resolution, making single enhanced data points hard to interpret. We note, however, that these enhancements are significantly higher in two of the Balmer lines and occur at a similar Doppler shift. The star exhibits a higher S/N than other SDSS spectra. It shows strong chromospheric emission lines that peak at approximately six times the flux of the surrounding continuum. The maximum Doppler velocity (manually read out from the spectrum) is approximately $600\: \mathrm{km~s^{-1}}$, $630\:\mathrm{km~s^{-1}}$, and $640\:\mathrm{km~s^{-1}}$ for $H\alpha$, $H\beta$, and $H\gamma$ respectively. A bulk velocity of $\sim 500\: \mathrm{km~s^{-1}}$ was derived using a Gaussian fit in the CME signature in $H\alpha$. Figure \ref{fig:cme_1_fitted} shows the fitted Gaussian profiles to the Balmer line core and the enhanced feature. The mass estimated from integrated flux measurements in $H\alpha$ using Eq. \ref{equ:houdebine:mass_cme} is approximately $3 \times 10^{17}\: \mathrm{g}$. The value is just an order of magnitude estimate and should be treated carefully because the method uses linear interpolation over the two data points to integrate over the enhanced range. The mass estimate based on the $H\gamma$ line is too uncertain because only one data point in the spectrum is enhanced.

While the quality of the plate is categorized as "bad" by SDSS, no data point flags in the wings of the emission lines indicate problematic data (only the $H\alpha$ line core is flagged in several spectra).

\begin{figure}
 \centering
 \resizebox{\hsize}{!}{\includegraphics{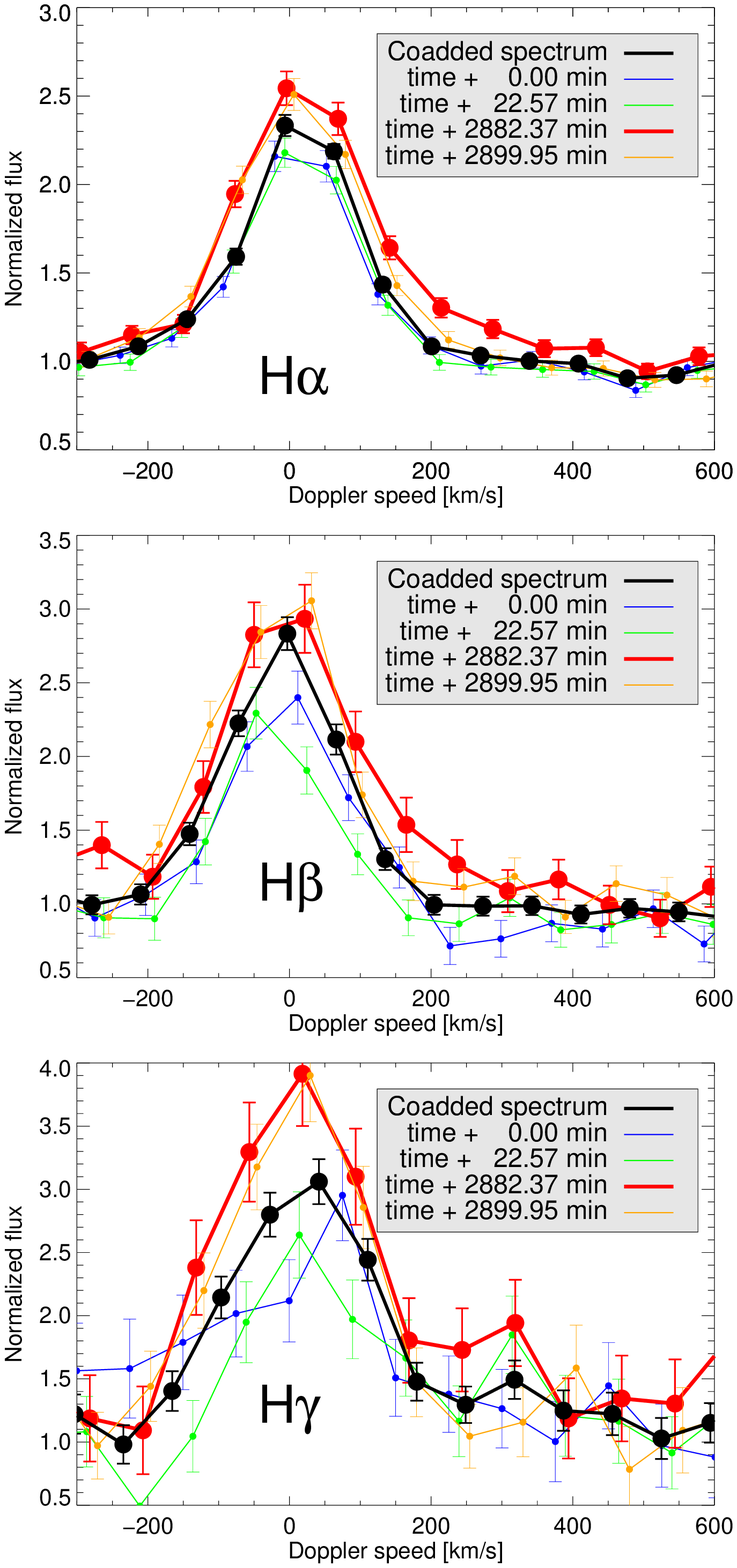}}
 \caption{CME candidate 2 in $H\alpha$ (upper panel), $H\beta$ (middle panel), and $H\gamma$ (lower panel). The spectral line at the time step showing the enhancement is marked in red.  \label{fig:cme_2}}
 \end{figure}

\begin{figure}
 \centering
 \resizebox{\hsize}{!}{\includegraphics{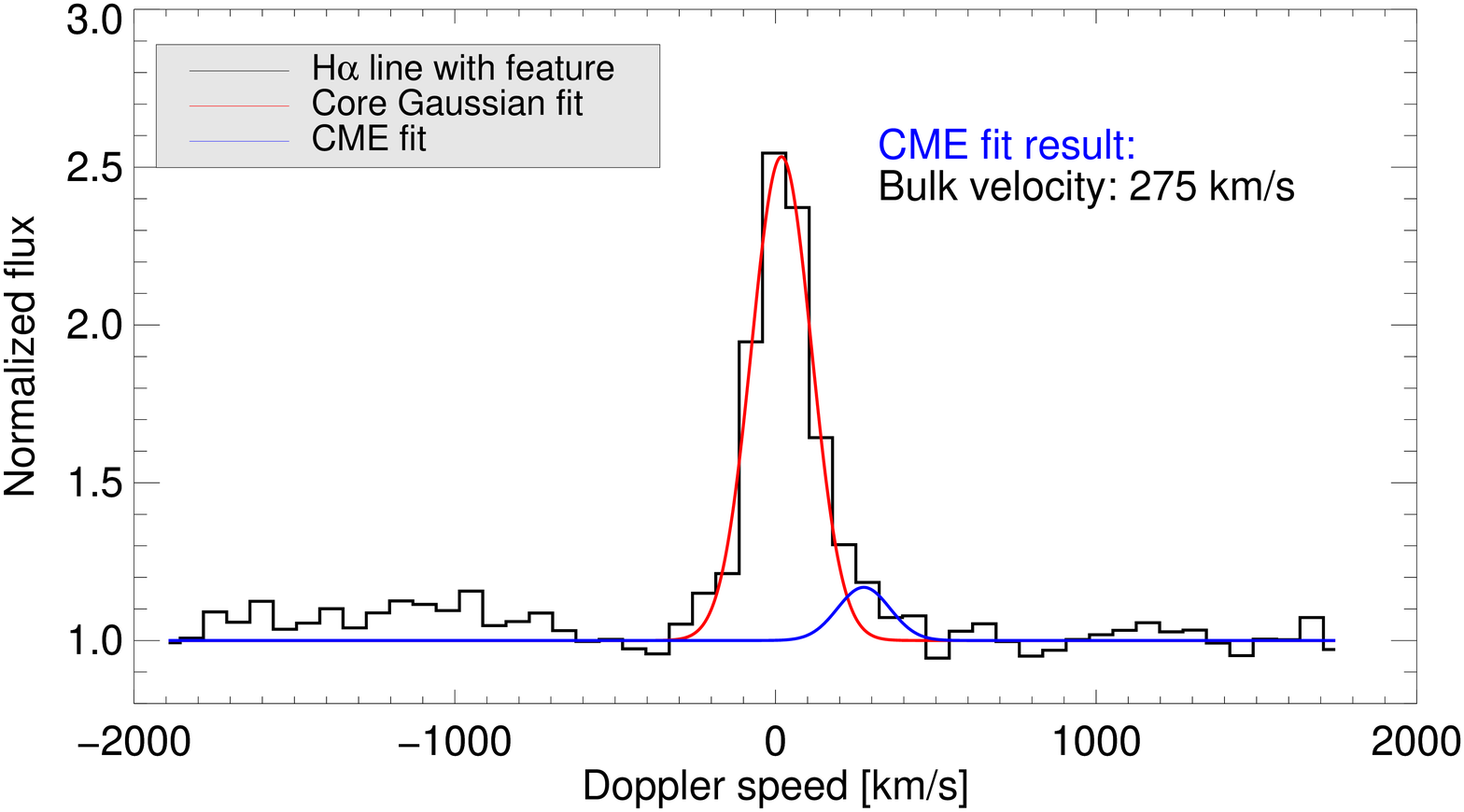}}
 \caption{$H\alpha$ spectral line of CME candidate 2. The line core (red line) as well as the possible CME feature (blue line) is fit; the bulk velocity of the Gaussian CME fit is annotated.  \label{fig:cme_2_fitted}}
 \end{figure}

\textbf{Candidate 2: SDSS J085556.66+174942.0 (Figs. \ref{fig:cme_2} and \ref{fig:cme_2_fitted}): }

The star with subtype M4e (SDSS classification M4) was found with the HTM algorithm while searching for flares. The flare is part of the M-star flare in the high S/N bin. The coadded spectrum consists of two sets, each containing two consecutive spectra. The latter two spectra (which are observed two days after the first set) show higher Balmer emission lines that are interpreted as flaring activity. Figure \ref{fig:cme_2} shows the spectra near the $H\alpha$, $H\beta$, and $H\gamma$ lines. Spectrum 3 shows a clear and significantly enhanced broad red wing in $H\alpha$ over seven neighboring data points. Two data points are enhanced in $H\beta$ and one data point in $H\gamma$, but none of them significantly. The noise in other chromospheric lines is too high to detect wing enhancements while the flare is still visible.

The maximum Doppler velocity derived is $710\: \mathrm{km~s^{-1}}$, $300\:\mathrm{km~s^{-1}}$, and $310\:\mathrm{km~s^{-1}}$ for $H\alpha$, $H\beta$, and $H\gamma,$ respectively. A bulk velocity of $\sim 300\: \mathrm{km~s^{-1}}$ is deduced, while the Gaussian fit to the CME signature in $H\alpha$ (see Fig. \ref{fig:cme_2_fitted}) did not work properly (because the normalization was not ideal on this wing). The mass estimated from integrated flux measurement in $H\alpha$ is $6.4 \times 10^{17}\: \mathrm{g}$.

The plate quality is categorized as "good" by SDSS. Four data point flags are present in the flaring Balmer emission line, which is not unusual for flaring activity.

\begin{figure}
 \centering
 \resizebox{\hsize}{!}{\includegraphics{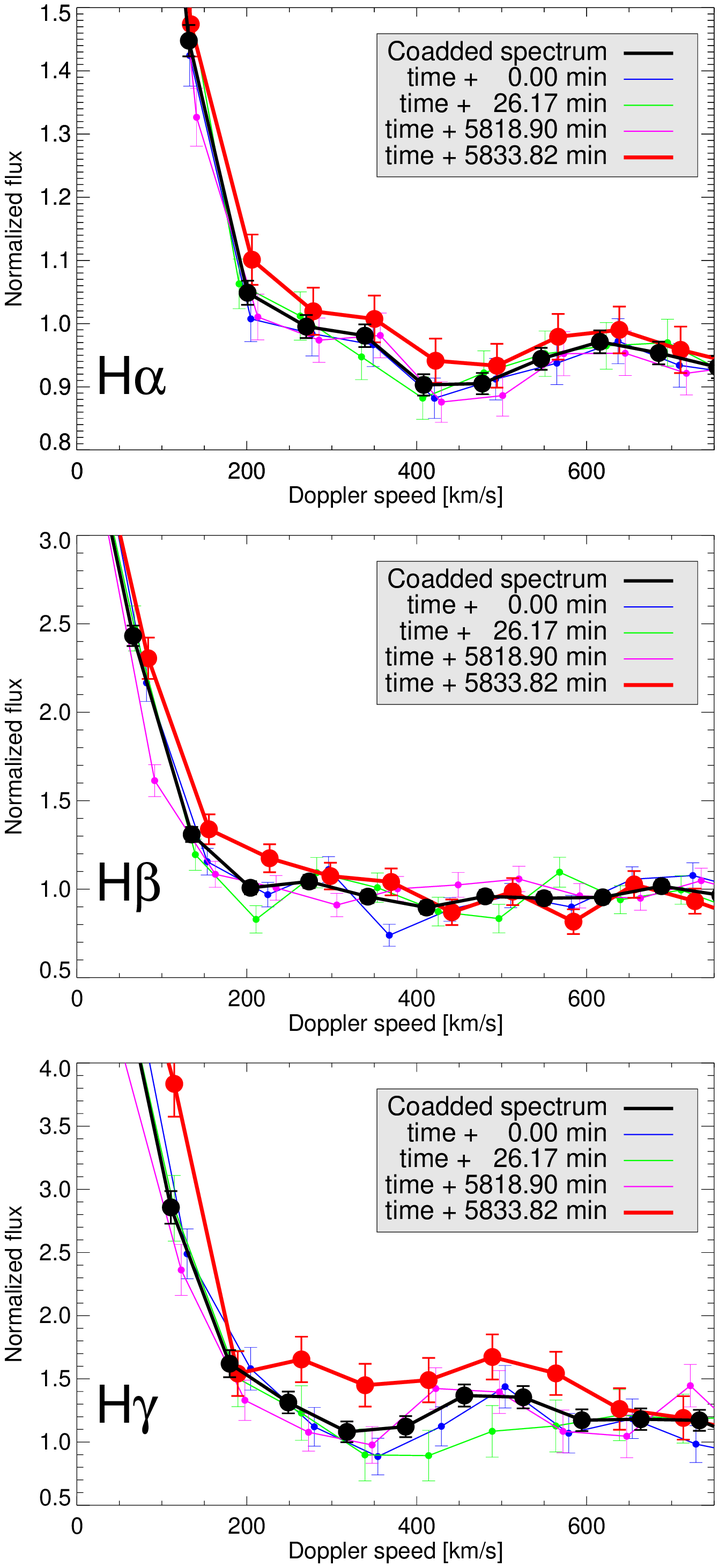}}
 \caption{Red wing of CME candidate 3 in $H\alpha$ (upper panel), $H\beta$ (middle panel), and $H\gamma$ (lower panel). The spectral line at the time step showing the enhancement is marked in red.}  \label{fig:cme_3}
 \end{figure}

\textbf{Candidate 3: SDSS J110743.30+075624.9 (Fig. \ref{fig:cme_3}):} 

The star with subtype M4e (SDSS classification M4) was found with the ELM algorithm. The coadded spectrum consists of two sets, each containing two consecutive spectra, with the latter set being observed four days after the first set. The exposure times are 20 minutes for the first two spectra, and 12 and 10 minutes for the latter two. Figure \ref{fig:cme_3} shows the red wing of the $H\alpha$, $H\beta$, and $H\gamma$ lines of these spectra. Spectrum 4 shows both an enhanced broad red wing in $H\alpha$ (between five and eight neighboring data points) and an enhanced blue wing (two neighboring data points). The enhanced data points in the red wing on their own are not significantly higher, that is, not above the error bars, but it is remarkable that there eight neighboring points are higher than all the other spectra in this range. The red $H\beta$ wing shows two enhanced data points (one significantly higher), and the red $H\gamma$ wing shows a clear enhancement of six higher points; three points are significantly enhanced. While it not significant considering the error bars, a red wing enhancement can also be seen in $H\delta$, the overlapping $H\epsilon / Ca\: II\: H$ line, and the $H8$ line (but none is visible in Ca II K). Because there appears to be an enhancement in all of these lines, the assumption that this is a CME becomes more probable, even though most of the data points are not significantly enhanced.

The maximum Doppler velocity is $670\: \mathrm{km~s^{-1}}$, $300\:\mathrm{km~s^{-1}}$, and $650\:\mathrm{km~s^{-1}}$ for $H\alpha$, $H\beta$, and $H\gamma$ respectively. A bulk velocity of $\sim 300\: \mathrm{km~s^{-1}}$ is deduced. Fitting a Gaussian to the CME signature in $H\alpha$ failed on the normalized flux of the spectrum with the enhanced wing. The mass estimated from the integrated flux measurement in $H\alpha$ and $H\gamma$ is $6 \times 10^{16}\: \mathrm{g}$ and $1.3 \times 10^{17}\: \mathrm{g}$, respectively. The integration over data points is possible in this case because of several neighboring data points in $H\alpha$ and $H\gamma$ (using linear interpolation).

The plate quality is categorized as "marginal" by the SDSS. No important data point flags affect this analysis, indicating no problem with the data.

\begin{figure}
 \centering
 \resizebox{\hsize}{!}{\includegraphics{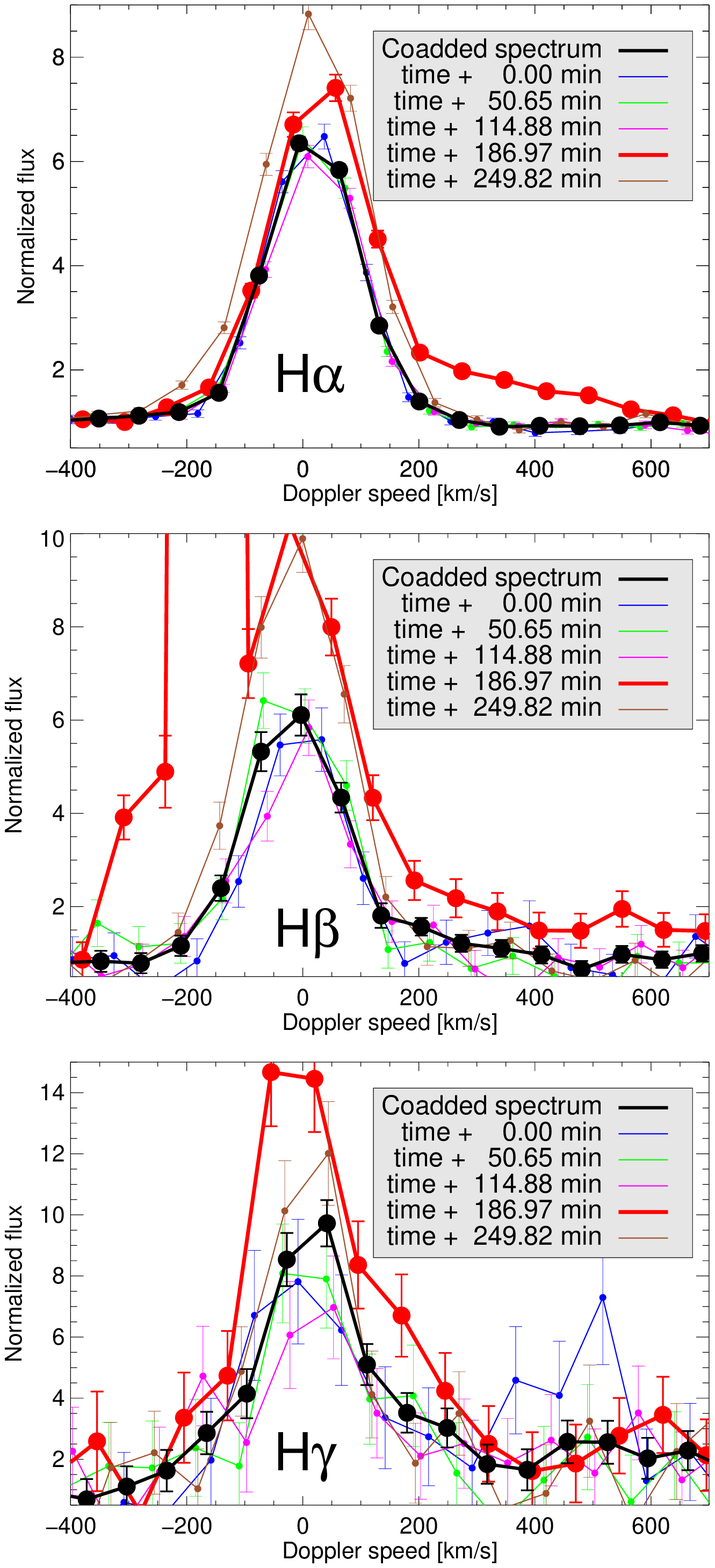}}
 \caption{CME candidate 4 in $H\alpha$ (upper panel), $H\beta$ (middle panel), and $H\gamma$ (lower panel). The spectral line at the time step showing the enhancement is marked in red.  \label{fig:cme_4}}
 \end{figure}

\begin{figure}
 \centering
 \resizebox{\hsize}{!}{\includegraphics{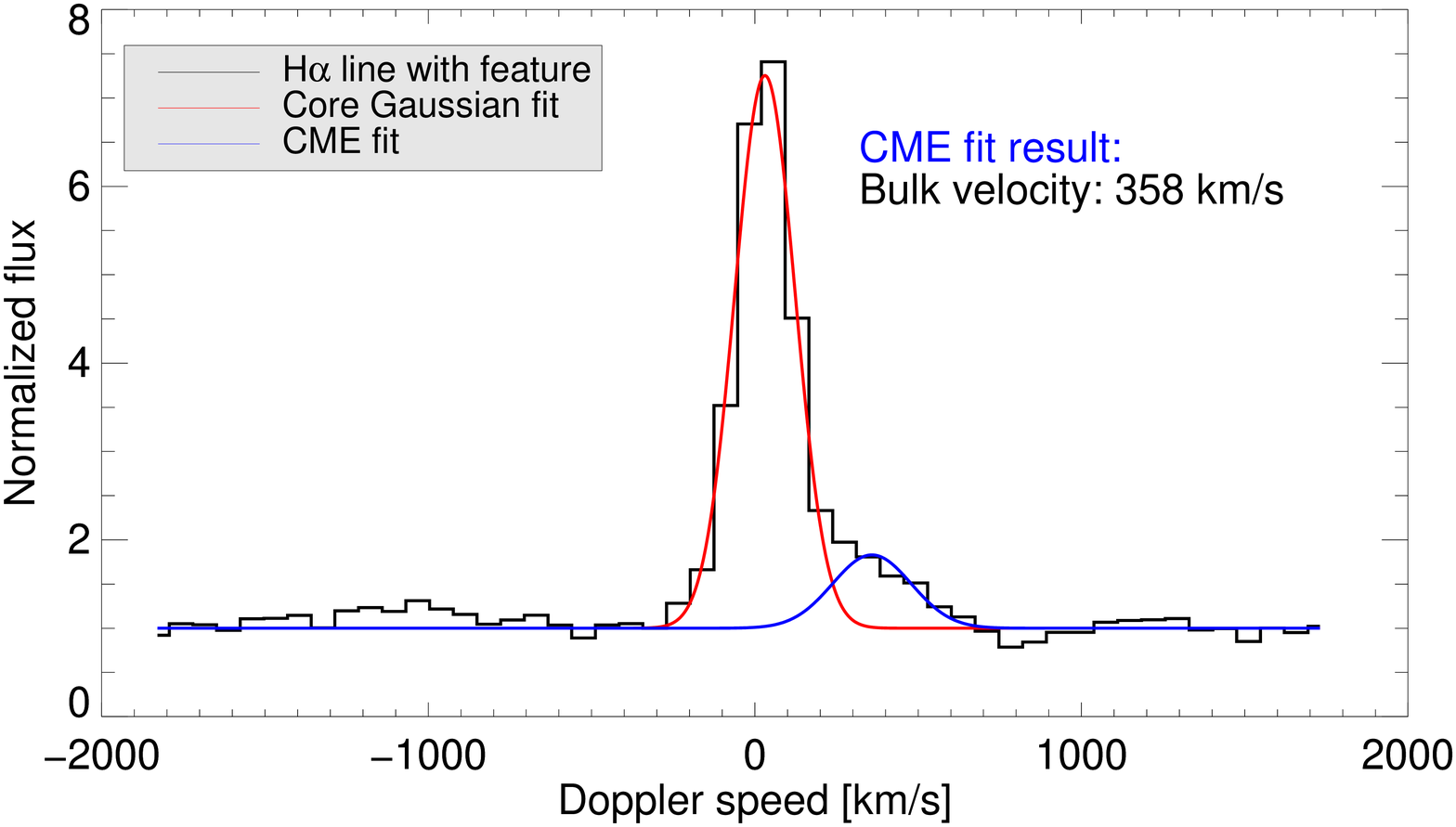}}
 \caption{$H\alpha$ spectral line of CME candidate 4. The line core (red line) as well as the possible CME feature (blue line) is fit; the bulk velocity of the Gaussian CME fit is annotated.  \label{fig:cme_4_fitted}}
 \end{figure}

\textbf{Candidate 4: SDSS J042139.64+264913.8 (Figs. \ref{fig:cme_4} and \ref{fig:cme_4_fitted}):}

The star with subtype M6 (SDSS classification M8) was found with the HTM algorithm when we searched for flares. The flare is part of the M-star flare in the medium S/N bin. The coadded spectrum consists of five single spectra observed within four hours. Figure \ref{fig:cme_4} shows the spectra near the $H\alpha$, $H\beta$, and $H\gamma$ lines. While the first three spectra show quiet Balmer emission lines (down to $H\gamma$), the fourth spectrum shows flaring activity with an enhanced peak and a broad red wing that is visible in the Balmer lines down to $H\delta$. The fifth and last spectrum is observed one hour after the last spectrum, only showing an enhanced emission line core (which is higher in $H\alpha$, of similar height in $H\beta,$ and somewhat lower than the previous one in $H\gamma$, but it is not visible in $H\delta$ at all). The red-wing enhancement shows seven significantly enhanced data points in $H\alpha$, eight  higher data points (three that are significantly enhanced) in $H\beta$, and two higher but not significantly enhanced points in $H\gamma$. An enhancement is also visible in $H\delta$, although it is not significantly higher. A cosmic-ray hit disturbs the blue wing of $H\beta$ in this spectrum.

The maximum Doppler velocity we derived is $700\: \mathrm{km~s^{-1}}$, $700\:\mathrm{km~s^{-1}}$, and $240\:\mathrm{km~s^{-1}}$ for $H\alpha$, $H\beta$, and $H\gamma,$ respectively. A bulk velocity of $\sim 360\: \mathrm{km~s^{-1}}$ was derived using a Gaussian fit to the CME signature in $H\alpha$. Figure \ref{fig:cme_4_fitted} shows the fitted Gaussian profiles to the Balmer line core and the enhanced feature. The mass estimated from integrated flux measurements in $H\alpha$ is $8.5 \times 10^{17}\: \mathrm{g}$. While the enhanced feature in $H\alpha$ is sufficiently broad to be safely integrated, the feature only shows one clearly enhanced data point in $H\gamma$, thus the mass estimation was omitted in $H\gamma$.

The plate quality is categorized as "bad" by the SDSS. The flaring $H\alpha$ and $H\beta$ lines are mostly flagged, as is the whole emission feature in the $H\alpha$ red wing. Because these values deviate strongly from the quiet spectra, the flagging of these values is not unexpected.

\begin{figure}
 \centering
 \resizebox{\hsize}{!}{\includegraphics{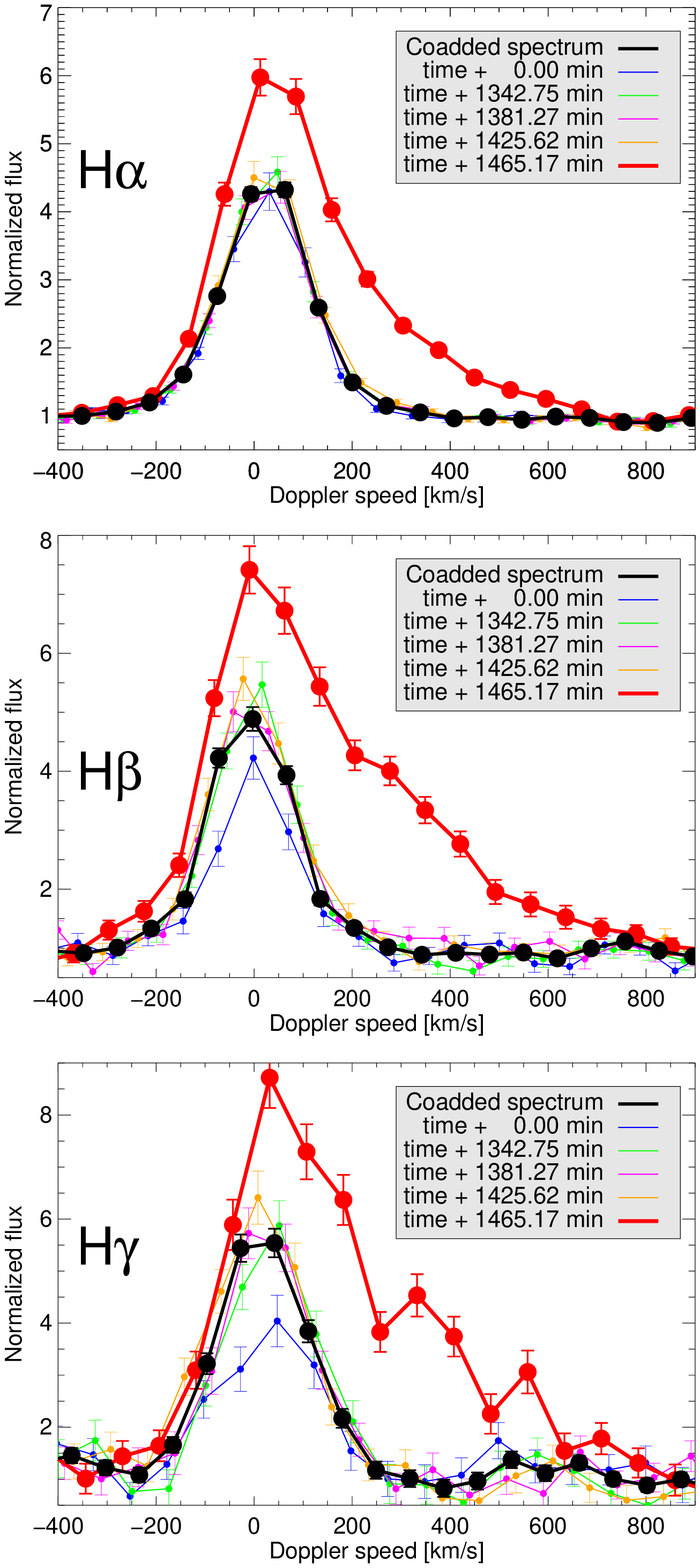}}
 \caption{CME candidate 5 (SDSS J053709.91-011050.3) in $H\alpha$ (upper panel), $H\beta$ (middle panel), and $H\gamma$ (lower panel). The spectral line at the time step showing the enhancement is marked in red.  \label{fig:cme_5}}
 \end{figure}

\begin{figure}
 \centering
 \resizebox{\hsize}{!}{\includegraphics{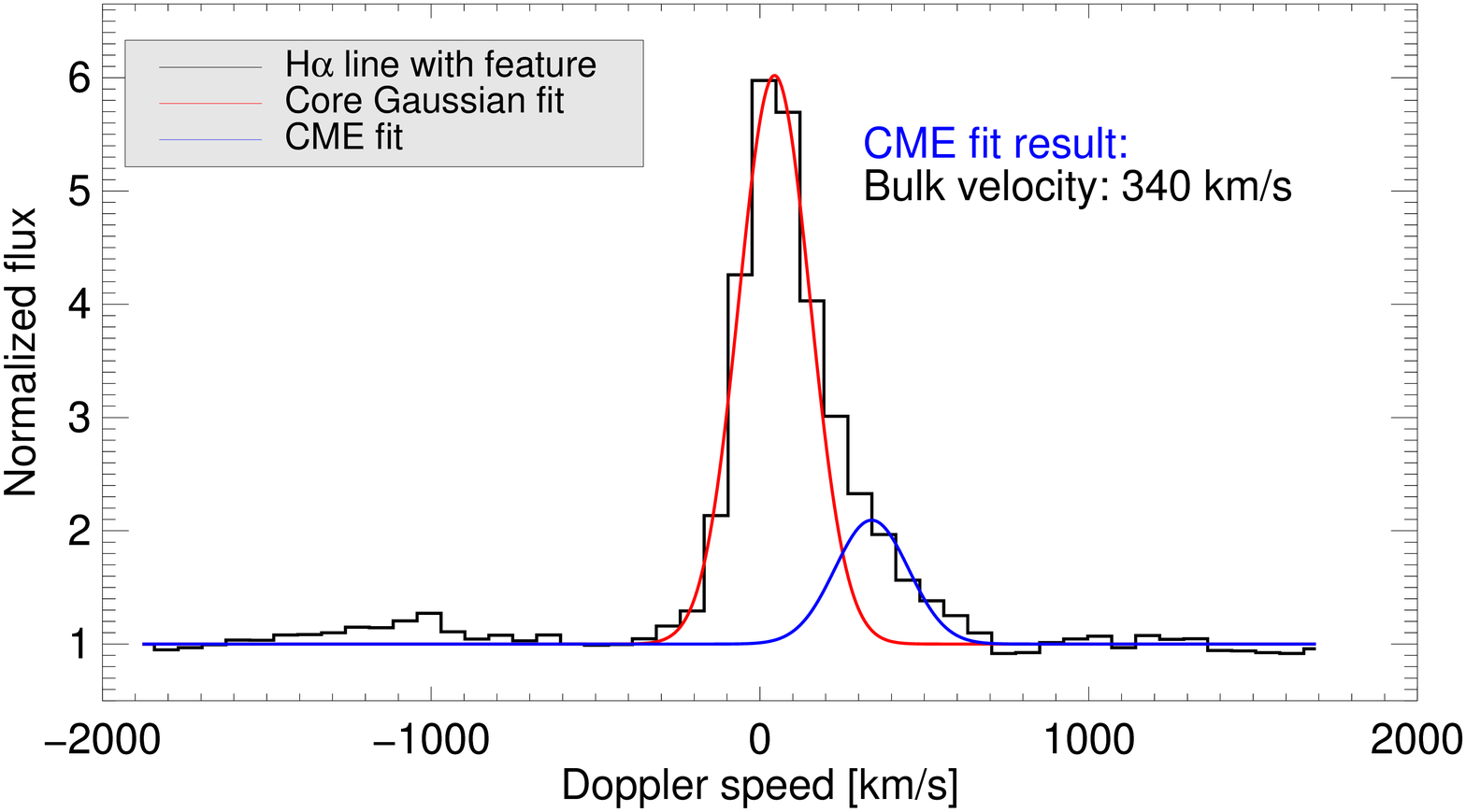}}
 \caption{$H\alpha$ spectral line of CME candidate 5. The line core (red line) as well as the possible CME feature (blue line) is fit, and the bulk velocity of the Gaussian CME fit is annotated.  \label{fig:cme_5_fitted}}
 \end{figure}

\textbf{Candidate 5: SDSS J053709.91-011050.3 (Figs. \ref{fig:cme_5} and \ref{fig:cme_5_fitted}):}

Candidate star 5 SDSS J053709.91-011050.3 is reported as a WTTS with SDSS classification M5 and was found with the HTM algorithm when we searched for flares. The flare is part of the M-star flare in the high S/N bin. The coadded spectrum consists of five single spectra, the first with an exposure time of 15 minutes, and the other four spectra build a consecutive time series one day after the first spectrum, also with 15 minutes exposure time each. Figure \ref{fig:cme_5} shows the spectra near the $H\alpha$, $H\beta$, and $H\gamma$ lines. The last spectrum shows a large flare in the chromospheric lines, which includes a broad flux enhancement in the red wing. The flare and flux excess is visible in all available Balmer lines down to $H\zeta$. The enhancement is also visible in the Ca II K line, but without the accompanying flare. The wing enhancement is significantly higher than the quiet spectrum in at least six neighboring data points in $H\alpha$, $H\beta$, and $H\gamma$ and is still significantly higher with at least one data point in $H\delta$, $H\epsilon$, and Ca II K.

The maximum Doppler velocity we derived is $710\: \mathrm{km~s^{-1}}$, $760\:\mathrm{km~s^{-1}}$, and $590\:\mathrm{km~s^{-1}}$ for $H\alpha$, $H\beta$, and $H\gamma,$ respectively. A bulk velocity of $\sim 340\: \mathrm{km~s^{-1}}$ was derived using a Gaussian fit to the CME signature in $H\alpha$. Figure \ref{fig:cme_5_fitted} shows the fitted Gaussian profiles to the Balmer line core and the enhanced feature. There is no mass estimate because no distance is available for the star.

The plate quality is categorized as "bad" by SDSS. Most data points of the $H\alpha$ emission line are flagged in all spectra, and the red wing excess is also flagged, which is expected because of the high difference of these values from the quiet spectrum.
 
\begin{figure}
 \centering
 \resizebox{\hsize}{!}{\includegraphics{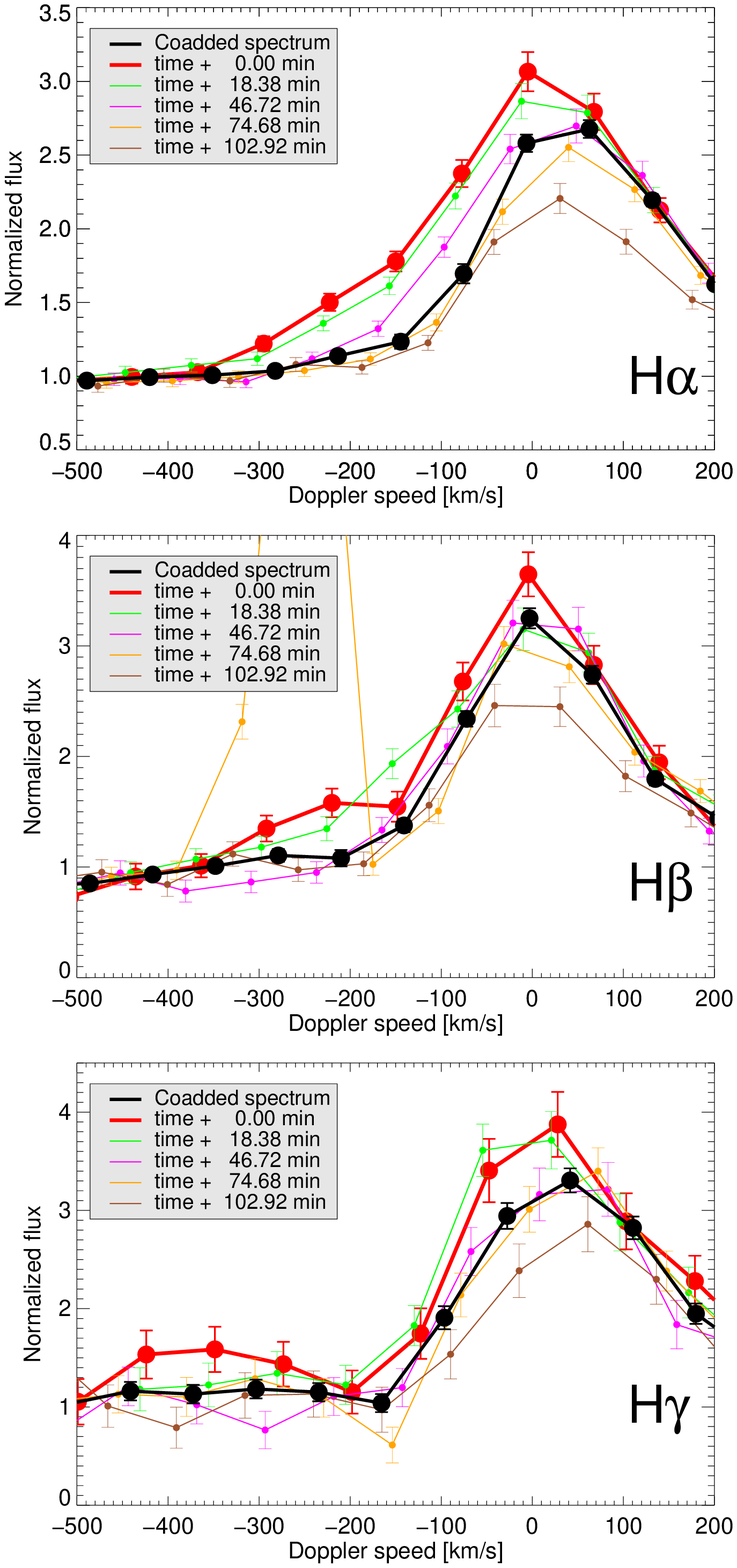}}
 \caption{Blue wing of CME candidate 6 (SDSS J052839.66-000322.5) in $H\alpha$ (upper panel), $H\beta$ (middle panel), and $H\gamma$ (lower panel). The spectral line at the time step showing the enhancement is marked in red.  \label{fig:cme_6}}
 \end{figure}

\begin{figure}
 \centering
 \resizebox{\hsize}{!}{\includegraphics{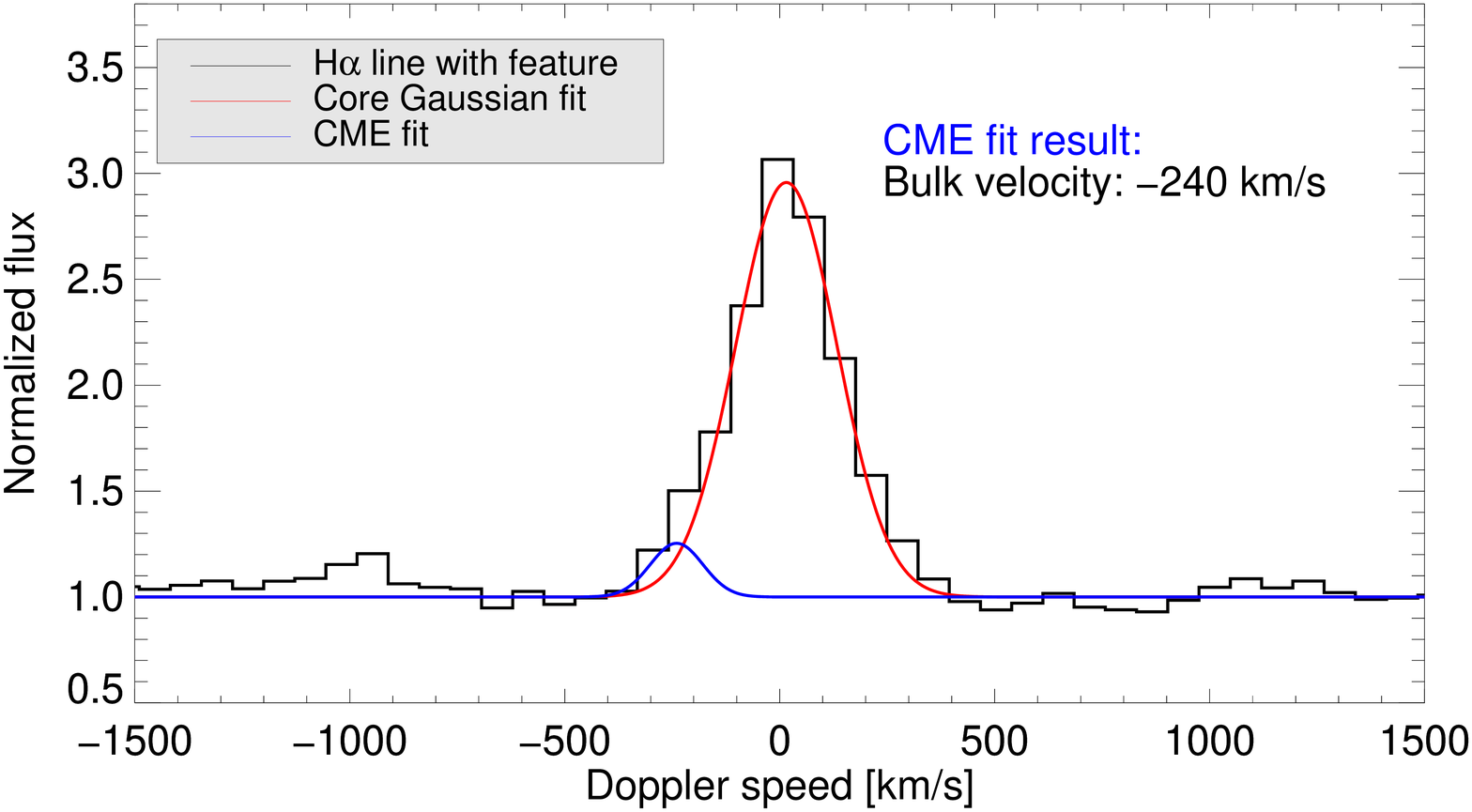}}
 \caption{$H\alpha$ spectral line of CME candidate 6. The line core (red line) as well as the possible CME feature (blue line) is fit, and the bulk velocity of the Gaussian CME fit is annotated.  \label{fig:cme_6_fitted}}
 \end{figure}

\textbf{Candidate 6: SDSS J052839.66-000322.5 (Figs. \ref{fig:cme_6} and \ref{fig:cme_6_fitted}):} 
Candidate star 6, SDSS J052839.66-000322.5, other designation "Kiso A-0903 163", is reported as an M4e type (SDSS classification M4) and was found with the HTM algorithm when we searched for flares. The flare belongs to the M-star flare in the high S/N bin. The coadded spectrum consists of five consecutive single spectra, with exposure times of 15 minutes for spectra 1 and 5, and 25 minutes for spectra 2, 3, and 4. Figure \ref{fig:cme_6} shows the spectra near the $H\alpha$, $H\beta$, and $H\gamma$ lines. The spectra show a decaying flare with a maximum in the first spectrum. The first three spectra also show an enhanced blue wing in $H\alpha$, which also decays over time. The entire blue wing (four  datapoints) of the first two spectra in $H\alpha$ is significantly enhanced compared to the last spectrum, which shows the least activity. The blue wing is also enhanced in $H\beta$, but only two data points in the blue wing in spectrum 1, and two are significantly higher than the quiescent spectrum. While an enhancement is still visible in $H\gamma$, no data point is significantly above the error bars of the quiet spectrum. The feature in $H\gamma$ appears to be at higher velocities than in the other Balmer lines. The enhancement also appears in $H\delta$, and two of the four enhanced data points are significantly higher.

The maximum Doppler velocity we derived is $-365\: \mathrm{km~s^{-1}}$, $-320\:\mathrm{km~s^{-1}}$, and $-450 \: \mathrm{km~s^{-1}}$ for $H\alpha$, $H\beta$, and $H\gamma,$ respectively. A bulk velocity of about $ -240\: \mathrm{km~s^{-1}}$ was derived using a Gaussian fit to the CME signature in $H\alpha$. Figure \ref{fig:cme_6_fitted} shows the fitted Gaussian profiles to the Balmer line core and the enhanced feature. The mass estimated from integrated flux measurements in $H\alpha$ using Eq. \ref{equ:houdebine:mass_cme} is $6.1 \times 10^{18}\: \mathrm{g}$. While the enhanced feature in $H\alpha$ is sufficiently broad to be safely integrated, the feature appears weaker in $H\gamma$, thus the $H\gamma$ mass estimation was omitted.

The plate quality is categorized as "bad" by SDSS. The $H\alpha$ line is flagged in every spectrum with the flag "BADSKYCHI". The blue-wing enhancements in the first three spectra are flagged with the flag "COMBINEREJ", which is a common flag for unusual enhancements. It states that these points are rejected for the coadded spectrum. 
 
%%%%%%%%%%%%%%%%%%%%%%DISCUSSION%%%%%%%%%%%%%%%%%%%%
\section{Discussion} \label{sec:discussion}

\subsection{Difference to Hilton et al. (2010)}\label{subdsubsec:diff_hilton}

\citet{hilton_2010AJ....140.1402H} performed a search for flares on M stars within an older data release of the spectroscopic SDSS data, finding 63 flaring events.  
While there is obvious overlap with our study, the differences in method and findings are substantial. Data release 14 that we used here covers additional and different objects compared to DR6, which was used by \citet{hilton_2010AJ....140.1402H}. \citet{hilton_2010AJ....140.1402H} preselected their sample using photometric SDSS flags and color selection to obtain M-star spectral types, while our selection of stars is solely based on the original SDSS spectral classification with restriction to F, G, K, and M dwarf or subgiant stars. \citet{hilton_2010AJ....140.1402H} made high and medium S/N samples and did not consider low S/N objects. Their quality cuts were based on the continuum level compared to the standard deviations. The emission line strength was measured by \citet{hilton_2010AJ....140.1402H} by defining a "flare line index" (FLI), where the mean value within the line region minus the continuum flux was divided by the continuum standard deviation. It measures the strength of the line and the significance compared to the surrounding noise. Their flare criteria were either based on difference in FLI values between maximum and minimum exposures in $H\alpha$ and $H\beta$ or on threshold FLI values in strong emission lines. The latter method used a null distribution of inactive stars to determine the quiet emission line value and declared a flare threshold for each subtype, each $H\alpha$ and $H\beta$ line, and each S/N sample based on false-discovery rate analysis. This method used by \citet{hilton_2010AJ....140.1402H} is quite different from the methods presented here, where Gaussian fits were used to solve the low S/N problem in the data. We also required temporal changes in the emission to be mandatory to declare a flaring state.

Of the 63 flares reported by \citet{hilton_2010AJ....140.1402H}, 41 ($65\:\%$) are part of the flares we found. Of the missing 22 flares, 8 were found by \citet{hilton_2010AJ....140.1402H} through their strong emission line criteria only, which explains that we did not detect them here because they showed too little or no amplitude changes to declare them as flaring based on our categorizations, even after visual inspection. Of the remaining 14 flares, one was found by the LTM and declared as too small for flaring condition. Two were found by the ELM search but were declared as nonflaring after visual inspection. One was found by the ELM search in the lowest S/N bin. Because of too many false detections, the objects detected by ELM in the lowest S/N bin were not analyzed further, therefore this flare was overlooked. Ten remaining flares were not detected by our algorithms. To understand why they were not detected, these ten objects are discussed in detail in Appendix \ref{sec:appendix_missing_flares}. The nondetections are caused by too weak changes in one Balmer line, missing data points, or by cosmic-ray hits in the reference spectrum that is used for the direct comparison.

\subsection{Constraints, statistics, and significance of detections}\label{subdsubsec:signif:findings_new}
The limited data quality allowed only a few flare detections compared to the size of the sample. The short and more or less random observational windows further complicated the search, and only allowed findings through the size of the  dataset.  
K and M stars (which have a higher probability to show flares) make up $24.1\: \%$  and $17.7\: \%$ of the stars in the original dataset. Only 9849 stars or $1.6\:\%$ of the original sample are M stars with high S/N, which is the subsample that showed the most flares (101 flares, which is $1\:\%$ of all high S/N M stars) and all CME wing enhancement candidates. Of all M stars in the dataset, $70\: \%$ have a very low S/N of $< 10$ near the $H\alpha$ line. These stars in the lowest S/N bin were still considered in the search. The significance of our flare findings in all S/N bins are reinforced by the (partly overlapping) flares reported by \citet{hilton_2010AJ....140.1402H} in the same dataset.

\subsection{Emission line stars in the flaring sample subset}\label{subsubsec:discussion:em_line_stars}
The sample of 281 flaring stars is clearly dominated by emission line stars, that is, by stars that show Balmer lines in emission even without a flare. The SDSS spectral classification does not explicitly state whether a star is an emission line star. Taking all stars where we found a source for the spectral type that explicitly state the emission line status (249 stars or $88.6 \%$ of all flaring stars), we find that $90 \%$ of these stars are declared to be emission line stars. After visual inspection, we find that no star in our resulting flaring list shows the $H\alpha$ line in absorption, even considering all nonflaring spectra. We find 14 stars (5\%) that showed neither emission nor absorption Balmer lines in quiescence, but significant Balmer emission lines during the flare. Ten of these 14 objects are part of the lowest S/N categorization.

Overall, these findings are expected: Balmer lines in emission are the indicator for magnetic activity, and our methods focused on fitting these lines (although both HTM and LTM should work on absorption lines as well). The low S/N of our data allowed us to only find flares on late-type stars with low luminosity and no flares on earlier types (F, G) that usually show Balmer lines in absorption.  

\subsection{Sample purity validation}\label{subsubsec:discussion:sample_purity}
Validating the purity of the luminosity type classification is an important step toward a robust flare detection statistic. Multicolor photometry, for example, can only determine the effective temperature, but not the luminosity class of an object and therefore cannot distinguish between close, faint dwarfs and distant, luminous red giant stars. 
\citet{mathur_red_giants_misclassified_2016ApJ...827...50M} performed asteroseismic analysis of 45400 stars, which were classified as dwarfs in the Kepler input catalog based on multicolor photometric classification. Mathur and collaborators found that $\sim$2$\: \%$ of the sample were cool giants. 

In light of this finding and because most spectral subtypes determined by the SDSS are given without luminosity class, it is sensible to confirm whether the detected flaring stars are indeed dwarf stars. Based on our analysis in Sect. \ref{subsec:flaring_occurrence,cmd,smaple_purity}, we find that only one of our 281 flaring stars appears to be a giant or subgiant K star.

\subsection{Flare energy and flaring occurrence}\label{subdsubsec:notes_energy}
The calculated flare energies cover a wide range from $3 \times 10^{28}$ erg to $2 \times 10^{33}$ erg. With a dataset this large, some findings of extraordinary strong flares (beyond $10^{32}\: \mathrm{erg}$ in $H\alpha$) are expected \citep[compare with stellar $H\alpha$ flare energies of approximately $10^{29} - 10^{30}$ erg,][]{hawley_flare_energies_adleo_2003ApJ...597..535H,leitzinger_flare_cme_search_blanco_2014MNRAS.443..898L}. Recent spectroscopy analysis of the active M dwarf AD Leo by \citet{muheki_spectroscopy_adleo_flare_2020A&A...637A..13M} derived similar results, with the strongest $H\alpha$ flare emitting an energy of $2.12 \times 10^{32}$erg.

The correlation between the energy or luminosity of the flares and the S/N of the objects is unclear from the whole dataset. This is a result of the mix of different spectral subtypes. Considering one specific subtype (e.g. M6 in Fig. \ref{fig:stellar_flare:_energy_subtype_snr}), we find that lower flare energies are found at higher S/N values (equivalent to stars that are closer to us), which is expected.

The lower threshold of flare energy compared to the distance (see Fig. \ref{fig:stellar_flare:_energy_distance}) was expected because only large flares are observable on distant stars. On the other hand, there is a visible upper threshold showing no high-energy flares within the surroundings of the Solar System. \citet{brasseur_galex_flares_2019ApJ...883...88B} found similar relations between flare energies and distances as an effect of observational sensitivity. The most probable explanations are selection effects in the data. With increasing distance, the volume and thus the number of stars and the probability of detecting high-energy flares obviously increases. We verified the distances per subtype and found that in the given dataset only few earlier subtypes (K stars and M0 or M1 stars) are close to the Sun (within 200 pc). This leads to missing detections of high-energy flares on nearby stars because energetic events occur mainly on earlier types.
Our findings that earlier types show higher maximum flare energies (see Fig. \ref{fig:stellar_flare:_energy_subtype_snr}) is consistent with previous reports in the literature (e.g. \citet{davenport_kepler_flares_2016ApJ...829...23D}, Fig. 5 or \citet{balona_flares_across_hrd_kepler_2015MNRAS.447.2714B}, Fig. 10).

The flaring occurrence rates shown in Fig. \ref{fig:flaring_percentage} are compatible with the result in \citet{kowalski_mdwarfs_sdss_stripe82_2009AJ....138..633K}, showing a rising flaring occurrence rate until spectral type M6. Unlike our result, which used a spectroscopic analysis of Balmer lines, \citet{kowalski_mdwarfs_sdss_stripe82_2009AJ....138..633K} used photometric light curves, giving the fraction of flaring epochs. \citet{west_m_dwarfs_sdss_spectral_types_distances_2011AJ....141...97W} determined the magnetic activity fraction for SDSS M dwarfs (based on "quiet" chromospheric emission) and reported the same relation. This further confirms our flaring occurrence results because flares are often related to high magnetic activity.

\subsection{Flare evolution}\label{subdsubsec:notes_flare_evol}
The higher number of rising flares compared to decaying flares given in Table \ref{tab:flare_evolution} is most probably an effect of the data and the search method we used. Our algorithm is most sensitive to strong changes in consecutive spectra. Decaying spectra may show too little changes to be detected due to the more gradual evolution. A second factor might be that the short rising period of flares \citep[expected to last less than 10 minutes, see, e.g., $H\alpha$ flare evolution curves in][]{kowalski_time_resolved_m_flares_2013ApJS..207...15K} may be shorter than the exposure times (up to 25 minutes) of the single spectra. Most of our rising flare cases appear to show two consecutive rising spectra (resulting in 20 -- 50 minutes), which is in contrast to the expected short timescale of the rise phase of flares. We could explain this as follows: the first flaring spectrum could show a mix of quiet and flare rise flux, including the peak. The integrated flux might therefore be lower than the subsequent spectrum, where the flux already decreases, but the integrated flux level is higher than the first spectrum. This would give the impression that the first and second flaring spectrum both increase and thus show the flare rise, although the flare is already in the decaying phase in the second spectrum. This would intrinsically favor rise phases over decay phases.

Some M star spectra show highly active Balmer lines with significant changes, sometimes showing several possible flares observed days apart from each other. The more prominent and easier to categorize flaring phase was chosen to represent the evolution of each set of spectra. The possibility of several flares in one set of spectra is thus not accounted for.

\subsection{CME candidates}\label{subdsubsec:cme_findings}
Of the six CME candidates showing Balmer line wing excesses presented in Sect. \ref{subsec_cme_candidates}, five show a red and one a blue wing enhancement. CME candidates 2 (Fig. \ref{fig:cme_2}), 4 (Fig. \ref{fig:cme_4}), and 5 (Fig. \ref{fig:cme_5}) show an enhanced red wing during a flare.
If this is a CME feature, it is counterintuitive to what is expected from the standard model of a CME onset because the flare together with the ejected material is directed toward the observer and therefore should be blueshifted. 
There are possible explanations for this feature: 
a) a backward-directed CME related to a flare occurring close to the limb; 
b) the visible feature is cool plasma material that falls back onto the surface of the star, albeit with very high velocities; c) chromospheric condensation \citep{kowalski_chrom_condensation_stars_2018ApJ...852...61K}, or d) a broad flare line profile superimposed with blue wing absorption from a CME, causing an apparent red-wing asymmetry. For CME candidate 2 (Fig. \ref{fig:cme_2}) in particular, we stress that the last spectrum no longer shows the red-wing enhancement, while the line core is still distinctly enhanced compared to the quiet spectra. A possible explanation in this case (other than downfalling material) is that the CME material becomes ionized on its way, making it invisible in Balmer lines in the last spectrum.

The CME candidate  6 (Fig. \ref{fig:cme_6}) shows a strong and broad wing enhancement during the flare similar to candidates 2, 4, and 5. The excess flux is blueshifted, however, which is in agreement with the interpretation as a CME.

The CME candidate 1 (Fig. \ref{fig:cme_1}) has significant red-wing enhancements in $H\alpha$, $H\beta,$ and $H\gamma$, but only with either one or two enhanced data points each, making the detection more doubtful. CME candidate 3 (Fig. \ref{fig:cme_3}), on the other hand, shows several enhanced data points in the red wing of each Balmer line, rendering it a good example of a possible CME in this dataset. Most of the enhanced data points for candidate 3 are not significantly higher than the other spectra, but the large number of neighboring and consistently higher data points and the appearance of the enhancement in eight chromospheric lines gives reason to take this event into consideration. Candidates 1 and 3 show no accompanying flare, which is in agreement with the theory of a backward-directed CME.

\subsection{CME properties, line asymmetries, rates, and nondetections }\label{subdsubsec:signif:cme_discussion}
The derived CME masses range from $6 \times 10^{16}\:$g to $6 \times 10^{18}\:$g. This lies within the expected range \citep[$10^{14}$ - $10^{19}\:$g, see Table 1 in ][]{stellar_cme_flare_relation_moschou_2019ApJ...877..105M}.  
\citet{odert_constraints_submitted_2019MNRAS} estimated the S/N ratios that are required to detect a certain CME mass depending on stellar spectral type. Their estimate for emission features \citep[see Fig. 1 in][]{odert_constraints_submitted_2019MNRAS} shows that for stars of type M5.5, at least an S/N of $\sim 20$ is required to detect CME masses above $6\times 10^{15}\:\mathrm{g}$ in $H\alpha$. For stars of type M2, the detectable CME masses with an S/N of $\sim 20$ rise to $\sim 5\times10^{16}\:\mathrm{g}$ in $H\alpha$. 
The S/N near $H\alpha$ of the identified CME candidates ranges between 17 and 28, and thus their derived masses lie well above the estimated minimum detectable value calculated by \citet{odert_constraints_submitted_2019MNRAS}.

The number of CME candidates is surprisingly low considering the minimum S/N - CME mass detection calculations by \citet{odert_constraints_submitted_2019MNRAS}, the assumed high association rate between high-energy flares and CMEs (based on solar relations), and the large number of high-activity late-type M stars. Still, this finding is consistent with other studies from literature. \citet{leitzinger_census_cmes_solar_like_2020MNRAS.493.4570L} used a similar approach by using optical spectral archive data and searched for CMEs on dF, dG and dK stars, finding no CME-like features and only little flaring activity on 425 stars with more than 3700 hours of observing time. \citet{muheki_spectroscopy_adleo_flare_2020A&A...637A..13M} analyzed $H\alpha$ line asymmetries in high-resolution spectroscopy of the active M star AD Leo with a total monitoring time of about 240 h. All determined Doppler-shift velocities of the study (blueshifts up to $260\: \mathrm{km~s^{-1}}$ and redshifts up to $190\: \mathrm{km~s^{-1}}$) lie well below the escape velocity of the star.

Explanations for our sparse findings based on the dataset quality may be a) low spectral resolution or b) flares giving wing enhancements that may outshine a possible CME-like feature. The long exposure times are another explanation for nondetection because the long integration times may diminish a short flux enhancement. Moreover, the minimum S/N - CME mass detection calculations by \citet{odert_constraints_submitted_2019MNRAS} are estimates for the maximum possible flux, further limiting the chances of detecting a CME, especially at low S/N.

A physical explanation for sparse CME findings may be that CMEs are suppressed by the strong overlying magnetic fields of the star, as was suggested by numerical simulations of stellar CMEs on M dwarfs by \citet[][]{alvarado_gomez_cme_suppression_2018ApJ...862...93A}. 
This would result in a significant reduction of velocity, suppressing weaker CMEs while only strong ejections could break free \citep{vida_quest_for_stellar_cmes_2019A&A...623A..49V}. Nonetheless, no clear and detached emission features were found. 
Another explanation lies in a possible strong ionization of the prominence material under the intensive EUV emission of active M stars \citep{howard_cme_ionisation_halpha_sun_2015ApJ...806..175H}. 

Alternative explanations for asymmetries arise by using the connection to flares \citep{canfield_sun_asymmetries_flares_1990ApJ...363..318C}. Balmer line asymmetries are generally associated with chromospheric material moving either downward or upward as a reaction to the impulsive energy input by flare-accelerated electrons causing chromospheric evaporation. The associated velocities to this mechanism on the Sun are usually in the range of tens of $\mathrm{km~s^{-1}}$ in chromospheric spectral lines, that is, far below the estimated velocities of our CME candidates \citep{li_chromosphere_evaporation_2019ApJ...879...30L}. \citet{cho_blue_asymmetry_sun_2016SoPh..291.2391C} reported strong preflare $H\alpha$ blueshifts on the Sun, with a wide velocity range of $-130\:\mathrm{km~s^{-1}}$ to $38\:\mathrm{km~s^{-1}}$, while the bulk velocity was low. The broadening was spatially correlated to a rising filament; however, there was no reported CME associated with this event. Using observations of a dMe 4.5 star, \citet{gunn_high_vel_chrom_evaportaion1994A&A...285..489G} reported a blue-wing emission feature similar to our CME candidates as a high-velocity chromospheric evaporation (maximum projected velocity of $600\: \mathrm{km~s^{-1}}$) during a flare, while the possibility of an associated CME was not discussed. Notably, their emission feature is also visible in the Ca II H \& K lines, similar to our CME candidate 5, and both show similar velocities and asymmetry shapes, but they have opposite directions.

Redshifted $H\alpha$ emission may arise from coronal rain (cooling plasma flows along post-flare loops) or from chromospheric condensation during the impulsive flare phase  \citep[][]{fuhrmeister_wing_asymmetries_exoplanets_around_mstars_2018A&A...615A..14F,vida_quest_for_stellar_cmes_2019A&A...623A..49V}.  Redshifts attributed to coronal rain in the solar case show velocities up to $120\:\mathrm{km~s^{-1}}$ with average values of roughly $60 - 70\:\mathrm{km~s^{-1}}$ \citep{antolin_coronal_rain_2012SoPh..280..457A,lacatus_solar_psot_flare_redshift_2017ApJ...842...15L,fuhrmeister_wing_asymmetries_exoplanets_around_mstars_2018A&A...615A..14F}. Chromospheric condensation arising from downward-moving material due to nonthermal electron beams is observed during the impulsive phases of solar flares with reported velocities of about $40\: \mathrm{km~s^{-1}}$ to $100\: \mathrm{km~s^{-1}}$ \citep{ichimoto_halpha_solar_flare_asymmetries_1984SoPh...93..105I}. More recent solar $H\alpha$ red-wing asymmetries arising from chromospheric condensation reported by \citet{asai_red_wing_solar_2012PASJ...64...20A} are at about $50\:\mathrm{km~s^{-1}}$. These redshifted velocities are significantly below the derived values of our CME candidates 1 - 5 with bulk velocities of $\sim 300\:\mathrm{km~s^{-1}}$ and maximum velocities up to $700\:\mathrm{km~s^{-1}}$. This might be an indicator that processes with higher velocities (such as CMEs) occur, although the measured bulk velocities are still below the escape velocities of the stars (which are about $500 - 600\:\mathrm{km~s^{-1}}$). We also recall that all CME velocity estimates are projected velocities, which decreases the true value by an unknown factor. 

Regarding the temporal evolution of our CME candidates, the low amount of spectra and long exposure times prohibit a detailed analysis. Still, CME candidates 2, 4, and 5 appear to show a red asymmetry during the first flaring spectrum, which might indicate chromospheric condensation. The velocities nevertheless significantly succeed chromospheric velocity values known from the Sun.  Recent efforts in numerical simulations of the coronal response during magnetically suppressed CME events on M stars suggest that weakly suppressed CMEs might display a temporal evolution of Doppler shifts from red to blue with velocities up to $200\:\mathrm{km~s^{-1}}$, while fully suppressed CMEs might show redshifted emission from in-falling material with speeds of $<-50\:\mathrm{km~s^{-1}}$ \citep[][]{alvarado_gomez_coronal_response_of_suppressed_cmes_2019ApJ...884L..13A}. However, we note that \citet[][]{alvarado_gomez_coronal_response_of_suppressed_cmes_2019ApJ...884L..13A} primarily focused on hot plasma and X-ray emission, in contrast to the cool plasma that produces Balmer lines. Alternatively, \citet{kuridze_line_asymmetries_gradient_2015ApJ...813..125K} showed that asymmetries in $H\alpha$ lines may also arise from shifts in the wavelength of the maximum opacity, in contrast to plasma motions.

This work did not employ an upper occurrence rate estimation on stellar CMEs due to the large fraction of spectra that are probably too low in S/N to perform a reasonable CME search. The overall observation time that is usable for CME analysis was therefore greatly reduced.

%%%%%%%%%%%%%%%%%%%%%%CONCLUSION%%%%%%%%%%%%%%%%%%%%
\section{Conclusions and outlook}\label{subsubsec:outlook}
We searched optical SDSS spectra of 630,162 stars (consisting mostly of F, G, K, and M dwarfs) in order to find flares and CMEs using both automatic algorithms and visual inspections. We found no flares on F and G type stars, 6 on K stars, 272 on M dwarfs, and 3 on M stars that are known to be WTT stars. We report 6 possible CME-like enhancements in the wings of Balmer lines and derive maximum velocities of up to 700 $\mathrm{km~s^{-1}}$ and CME masses between $6 \times 10^{16}\:$g and $6 \times 10^{18}\:$g.

With stellar CMEs playing a crucial role in the evolution of exoplanetary atmospheres and with their effect on stellar mass and angular momentum loss, it is important to constrain CME occurrences and parameters \citep{lammer_cme_mstars_impact_hability_2007AsBio...7..185L,airapetian_space_waether_impact_2020IJAsB..19..136A}. Like \citet{hilton_2010AJ....140.1402H}, we demonstrated that searching through large archival data proves a viable method to search for stellar flares in optical spectra, even with limited S/N. A search for associated CMEs is reasonable considering minimum S/N constraints for CME masses \citep{odert_constraints_submitted_2019MNRAS} and the high association rate between energetic flares and CMEs on the Sun \citep{solar_cme_flare_correlation_yashiro_2009IAUS..257..233Y}. While other works have implemented similar search approaches \citep[e.g.,][]{vida_quest_for_stellar_cmes_2019A&A...623A..49V,leitzinger_census_cmes_solar_like_2020MNRAS.493.4570L}, numerous data archives with spectral data have not yet been analyzed within the scope of stellar CME search.

Still, considerable information is lost in similar data archives because of the short observational time windows (and thus only few single spectra) for each star, compared to long and continuous observations of one single star. To solve the problem of blocking a telescope for long observations, a flare-alert system like the one presented by \citet{hanslmeier_flare_alert_2017CEAB...41...67H} might be a promising concept. When continuous photometric observations with small telescopes are used, a detected flare can be reported in real time to larger observatories so that they can conduct spectroscopic follow-up observations.
While this work focused on main-sequence stars, we may shift the focus to include pre-main-sequence stars in order to search for CMEs because these objects are known for their high activity.

We focused on CME-like features that are visible at all Balmer lines, while the possibility of CME material being optically thick in one Balmer line only might give reason to reconsider this approach. This needs to be further analyzed.

%%%%%%%%%%%%%%%ACKNOWLDEGEMENTS%%%%%%%%%%%%%%%%%%%
\begin{acknowledgements}
F.K., M.L., and P.O. gratefully acknowledge the Austrian Science Fund (FWF): P30949-N36 for supporting this project. M.L., P.O., and A.M.V. gratefully acknowledge support by the Austrian Space Applications Programme of the Austrian Research Promotion Agency FFG (ASAP-14 865972, BMVIT). Funding for the Sloan Digital Sky Survey (SDSS) has been provided by the Alfred P. Sloan Foundation, the Participating Institutions, the National Aeronautics and Space Administration, the National Science Foundation, the U.S. Department of Energy, the Japanese Monbukagakusho, and the Max Planck Society. The SDSS Web site is \url{http://www.sdss.org/}. The SDSS is managed by the Astrophysical Research Consortium (ARC) for the Participating Institutions. The Participating Institutions are The University of Chicago, Fermilab, the Institute for Advanced Study, the Japan Participation Group, The Johns Hopkins University, Los Alamos National Laboratory, the Max-Planck-Institute for Astronomy (MPIA), the Max-Planck-Institute for Astrophysics (MPA), New Mexico State University, University of Pittsburgh, Princeton University, the United States Naval Observatory, and the University of Washington. This research has made use of the VizieR catalogue access tool, CDS, Strasbourg, France (DOI : 10.26093/cds/vizier). The original description  of the VizieR service was published in 2000, A\&AS 143, 23. This work has made use of data from the European Space Agency (ESA) mission Gaia (https://www.cosmos.esa.int/gaia), processed by the Gaia Data Processing and Analysis Consortium (DPAC, https://www.cosmos.esa.int/web/gaia/dpac/consortium). Funding for the
DPAC has been provided by national institutions, in particular the institutions participating in the Gaia Multilateral Agreement.
\end{acknowledgements}

%%%%%%%%%%%%%BIBLIOGRAPHY%%%%%%%%%%%%%%%%
\bibliographystyle{aa} % style aa.bst
\bibliography{mybib} % your references Yourfile.bib

%%%%%%%%%%%%%%%%%%%%APPENDIX%%%%%%%%%%%%%%%%%%%%%%
\appendix

%%%%%%%%%-A-%%%%%%%%%%%%
\section{Missing flares of Hilton et al. (2010)}\label{sec:appendix_missing_flares}

Using notation from \citet{hilton_2010AJ....140.1402H}, Table 3, we identified the stars by their right ascension and declination coordinates. We list them below

\textbf{Star R.A. 355.347630 and Decl. -0.646637:}
This M1 star in the high S/N bin probably showed too little variation in the $H\alpha$ line height for detection. It was probably not found by the high emission line algorithm due to the high continuum flux compared to the $H\alpha$ line height.

\textbf{Star R.A. 128.514690 and Decl. 23.718951:}
This M3 star in the high S/N bin shows flaring activity in later Balmer lines, but not in $H\alpha$.

\textbf{Star R.A. 134.892240 and Decl. 28.224011:}
This M3 star in the high S/N bin shows small variation (decay) in one spectrum. It was probably not detected due to too small temporal changes in $H\beta$.

\textbf{Star R.A. 154.133280 and Decl. 36.066472:}
This M3 star in the high S/N bin shows slowly decaying emission lines. The changes in $H\alpha$ are most likely too small for a detection.

\textbf{Star R.A. 179.329230 and Decl. 36.899333:}
This M3 star in the high S/N bin shows too little changes in $H\alpha$. The spectrum showing the highest differences in the in $H\alpha$ emission line does not show the same behavior in $H\beta$.

\textbf{Star R.A. 8.116274 and Decl. -0.669970:}
This M4 star in the high S/N bin shows a significantly high rising flare. There was probably no detection due to a cosmic-ray hit near the peaking spectrum in $H\beta$, which caused the Gaussian fit to fail.

\textbf{Star R.A. 123.544780 and Decl. 7.833261:}
This M4 star in the medium S/N bin shows variation of the $H\alpha$ and $H\beta$ lines, but too little to be detected with our algorithms. The change from one time instant to another between flaring spectra in $H\beta$ appears to be too small.

\textbf{Star R.A. 237.126050 and Decl. 51.509009:}
This M4 star in the high S/N bin shows a huge flare visible in all chromospheric lines. The flare was not detected due to missing data points in the wing of the $H\alpha$ line.

\textbf{Star R.A. 329.517010 and Decl. -8.355448:}
This M6 star in the low S/N bin shows its peak Balmer line enhancement in its last spectrum. It was probably not detected becaiuse the previous spectrum in $H\beta$ showed a cosmic-ray hit, giving rise to a faulty comparison between the flaring and the previous spectra. 

\textbf{Star R.A. 182.070410 and Decl. 8.757879:}
This M9 star in the lowest S/N bin shows a clear flare in the Balmer lines in its last spectrum. It was probably not detected due to a cosmic-ray hit near the Balmer line in the previous spectrum in $H\alpha$.

%%%%%%%%%%%%%%%%%%%%%%%-B-%%%%%%%%%%%%%%%%%%%
\section{Special cases in the flaring list}\label{sec:appendix_special_cases_flares}
\textbf{M stars:}

Several stars were identified to be cataclysmic variables, showing extreme shifts in the spectra and chromospheric line height variations. Together with other, possible close ( and eclipsing) white dwarf + late-type binaries with extreme line shifts, they were excluded from the list.

After further inspection, several stars still present in the list are also possible WD+dM binaries. One object (SDSS J045325.65-054459.2) was found to be most likely a late-type M star in front of a more distant star of an earlier type. This becomes clear from the available GAIA distances, which reveal two sources within two arcseconds. \citet{rebassa_wdbinaries_ind_sdss_2013MNRAS.433.3398R} came to the same conclusion, stating the object as a superposition of two main-sequence stars.

Other stars given as possible binaries were still included if there was no sign of significant shift or unusual behavior of the lines, which would otherwise indicate close binaries. The object SDSS J005506.77-005702.4 is given as either a detached or (with low confidence) semidetached star by \citet{becker_perdiodic_variability_low_mass_stars_2011ApJ...731...17B}. With only three consecutive spectra, the possible shift due to being a binary system in the data is too low to make concrete assumptions; the object is therefore kept in the list with the flaring attribute "probable".

Object SDSS J141528.99+123127.4 was found to be in front of a galaxy. Manual inspection revealed that the GAIA DR2 parallax closest to the stellar coordinates obtained by cross-matching gave the value of the galaxy (parallax around 0 mas). The distance was replaced with the real stellar distance value. Object SDSS J161237.41+164556.0 also appears to be in front of another object, resulting in two GAIA DR2 distances. Both distances are similar \citep[225 versus 256 parsec,][]{bailer-jones_distances_gaia_dr22018AJ....156...58B}.

The spectrum of object SDSS J162027.54+364002.8 consists of two stars (M4 and M3) that lie too close to each other to be resolvable as independent sources because the entire diameter in SDSS plates is larger than their angular distance. The distances given by GAIA DR2 are 300 and 350 parsecs, respectively.

\textbf{K stars:} 

The star SDSS J161924.09+061554.2, marked as subtype K3 in SDSS, had a flare that changed its Balmer absorption lines into emission lines. It was identified to be an RS Canum venaticorum star \citep{drake_periodic_variable_catalogue2014ApJS..213....9D}.

The star SDSS J011707.01+250359.8, marked as subtype K3 in SDSS, showed a new emission line in $H\alpha$ and $H\beta$ at one time instance. It was identified to be a eclipsing binary of type W UMa \citep{drake_periodic_variable_catalogue2014ApJS..213....9D}.

The star SDSS J075835.12+482523.6 is marked as a K5 star in the SDSS and shows high chromospheric emission lines with variation over time. The spectra show clear shifts in the data, which is an indication of a close binary object. This assumption is confirmed by \citet{kleinmann_sdss_dr7_wd_2013ApJS..204....5K}, who declared it a white dwarf+M0e binary, and by \citet{watson_varibale_star_index_2006SASS...25...47W}, who recognized the object as a cataclysmic variable.

The star SDSS J215455.17+011414.5, marked as subtype K5 in the SDSS, shows significantly higher emission lines in $H\alpha$, $H\beta$, and $H\gamma$ and especially high Ca H and K emission lines at two time instances. The spectrum as a whole is redshifted by $\sim2\:\text{\AA}$ in these instances. Even without the flare, the star showed these lines in emission. \citet{segue_giant_survey_2014ApJ...784..170X} determined the star to be a K giant star on the RGB, while \citet{kowalski_mdwarfs_sdss_stripe82_2009AJ....138..633K} in their search for flaring M dwarfs used the spectral classification dM0. When data from GAIA are used, the star appears to be on the upper part of the main sequence, while Gaia data also give information on the radius ($2.1972 \mathrm{R_{\odot}}$) and luminosity ($2.153 L_{\odot}$) (GAIA DR2). This clearly suggests that the star is not an M dwarf star. Figure \ref{fig:sample_purity} confirms this assumption because it is the only flaring star above the Sun-like star line. We kept the flare in the final flaring list as a possible K giant or subgiant. The high emission lines might also indicate a T Tauri type, but there is no literature result to further strengthen this claim. 

The star SDSS J000500.22-043228.8 is marked as a K5 star in the SDSS, while an unpublished work of Skiff et al.(2013) gave it the spectral type M0Ve. \citet{heinze_atlas_variable_stars_catalogue_2018AJ....156..241H} declared the object a distant binary. The spectral type K5 appears to be more accurate according to the stellar position in Fig. \ref{fig:sample_purity}. The SDSS spectra show a possible flare evolution on both $H\alpha$ and $H\gamma$. The whole spectrum is shifted over time, which is an indicator for binaries. Flags in the red spectrum range and missing Balmer lines of a higher order resulted in the "probable" flare categorization of the event.

\end{document}